\def\re#1{(\ref{#1})}
\def\beq{\begin{equation}}
\def\eeq{\end{equation}}
\def\beeq{\begin{eqnarray}}
\def\beeqn{\begin{eqnarray*}}
\def\eeeq{\end{eqnarray}}
\def\eeeqn{\end{eqnarray*}}
\def\nome#1{{\label{#1}}}

\def\a{\alpha}

                  \def\G{\Gamma}
\def\de{\delta}                 
\def\e{\varepsilon}

\def\m{\mu}
\def\n{\nu}

\def\s{\sigma}                  
\def\th{\theta}

\newcommand{\DD}{{\cal D}}

\newcommand{\GG}{{\cal G}}

\newcommand{\OO}{{\cal O}}

\newcommand{\ZZ}{{\cal Z}}
\newcommand{\WW}{{\cal W}}
\newcommand{\zk}{{\cal Z}_{k}}
\newcommand{\wk}{{\cal W}_{k}}
\newcommand{\wka}{{\cal W}_{k,a}}

\newcommand{\lp}{\left(}
\newcommand{\rp}{\right)}
\renewcommand{\lq}{\left[}
\renewcommand{\rq}{\right]}
\newcommand{\lgr}{\left\{}
\newcommand{\rgr}{\right\}}
\newcommand{\identity}{1\hspace{-0.4em}1}
\newcommand{\no}{\nonumber}
\newcommand{\ph}{\phantom} 
\def\abs#1{\left|#1\right|}
\def\tr{\,\mbox{Tr}\,}
\def\frac#1#2{ {{#1} \over {#2} }}

\def\half{\mbox{\small $\frac{1}{2}$}}

\def\ie{\hbox{\it i.e.}{ }}

\def\bom#1{\mbox{\boldmath$#1$}}

\def\kap{\bom{k}}

\documentstyle[preprint,aps,floats,amssymb]{revtex}
\begin{document}
\input{epsf}
\draft
\newfont{\form}{cmss10}

\title{Wilson loops in the adjoint representation and 
multiple vacua in two-dimensional Yang-Mills theory 
}

\author{A. Bassetto$^1$, L. Griguolo$^2$ and
F. Vian$^1$} 
\address{$^1$Dipartimento di Fisica "G. Galilei" and
INFN, Sezione di Padova,\\
Via Marzolo 8, 35131 Padua, Italy\\
$^2$Dipartimento di Fisica ``M. Melloni'' and
INFN, Gruppo Collegato di Parma, \\ Viale delle Scienze, 43100 Parma, Italy}
\maketitle
\begin{abstract}
                      
\noindent
$QCD_2$ with fermions in the adjoint representation is invariant under
$SU(N)/Z_N$ and thereby is endowed with a non-trivial vacuum structure
($k$-sectors). The static potential between adjoint charges, in the 
limit of infinite mass, can be therefore obtained by computing Wilson 
loops in the pure Yang-Mills theory  with the same non-trivial 
structure. When the (Euclidean) space-time is compactified on a sphere 
$S^2$, Wilson loops can be exactly expressed in terms of an infinite 
series of topological excitations (instantons). The presence of 
$k$-sectors modifies
the energy spectrum of the theory and its instanton content. For the
exact solution, in the limit in which the sphere is decompactified,
a $k$-sector can be mimicked by the presence of $k$-fundamental charges at
$\infty$, according to a Witten's suggestion. However, this property neither
holds before decompactification nor for the genuine perturbative solution 
which corresponds to the zero-instanton contribution on $S^2$.  
\end{abstract}
\vskip 2.0truecm
DFPD 00/TH 10

\noindent
UPRF-00-01


\vskip 3.0truecm
\vfill\eject

\narrowtext
\section{Introduction}

\noindent
One of the fundamental problems in discussing the quantum dynamics of 
nonabelian gauge theories is to understand their vacuum structure. 
Many important aspects of four-dimensional QCD such as chiral symmetry 
breaking and confinement, are believed to be related to some non-trivial 
properties of the vacuum itself. Particularly intriguing in this context is 
the appearance of a topological parameter, the $\theta$ angle \cite{Ja76}, 
directly related to the existence of multiple vacuum states and to 
quantum tunnelling between them mediated by instanton effects \cite{Ca78}. It 
would be extremely interesting to understand how physical quantities could 
depend on $\theta$, but, unfortunately, four-dimensional QCD appears too 
complicated to deal with. A non-perturbative analysis is in fact 
mandatory, and satisfactory results concerning the $\theta$-dependence 
have been up to now obtained only in particular supersymmetric cases
\cite{shif}. 

Nevertheless, a formulation of the Quantum Field Theory exists, 
in which, at least in principle, 
the exact ground state is described as a simple Fock vacuum: the Light-Front 
(LF) Quantum Field Theory, where the theory is quantized not on a space-like 
surface but on a light-like one (see \cite{light} for an extensive review). 
Actually,
in four-dimensional gauge theories such an approach is far from being 
simple, due to the intricate dynamics of the so-called zero-modes that mix 
with the Fock vacuum and substantially spoil the kinematical character of 
the ground state \cite{Bur}. This problem does not exist in two dimensions
(zero-modes are related to transverse degrees of freedom in LF quantization) 
and recently it has 
been shown \cite{capo} that, at the perturbative level,  
LF quantization encodes a
complicated instanton dynamics present in the equal-time (ET) formulation. 
Therefore two-dimensional gauge theories candidate  themselves as the 
simplest models in which the influence of 
topological parameters on physical quantities and their interplay with the 
vacuum structure, as well as on their perturbative (if any)
interpretation, can be probed.  
The best example of a theory which admits multiple vacua and shares 
relevant features with four-dimensional gauge theories is $QCD_2$
with  adjoint 
fermions, as noticed many years ago by Witten \cite{wittheta}: here
a single 
integer $k$ labels inequivalent vacua, taking 
the value $0,1,..,N-1$ in the $SU(N)$ case. This model exhibits a 
rich spectrum \cite{Add}, a Hagedorn transition at finite temperature 
\cite{Ko} 
and interesting confining vs screening properties \cite{forr}: 
the presence of adjoint 
matter in some sense mimics
transverse degrees of freedom inducing a complex 
behaviour. Of particular interest is the link between topological sectors, 
multiple vacua, instantons and condensates: the presence of a fermionic 
condensate, in the $SU(N)$ case, can be argued
by looking at the 
bosonized version of the theory \cite{smilga2} and it is also expected from arguments 
of quark-hadron duality \cite{Ko}. Nevertheless, only for $N=2$ the instanton 
computation, relating the presence of zero-modes of the Dirac operator in a 
non-trivial background to the condensate, produces the desired result
\cite{smilga1}.  
In the general case no satisfactory answer is known to our knowledge,
since on one hand the results obtained within the ``Discretized Light-Cone
Quantization'' (DLCQ) depend on the size of the discretization parameter 
\cite{pin}, on the other hand the approach of \cite{lenz} is
essentially restricted to the $SU(2)$ case in the small volume limit.
 
We therefore find  worthwhile  examining these problems at the light
of the 
result of \cite{capo,noi}, where LF quantization of Yang-Mills theory has been shown 
to reproduce  the non-perturbative series of instantons that 
naturally appears in the ET formulation in a simple manner.
Our present
goal is to investigate   carefully  the vacuum properties in both quantization
schemes
and to understand the relation with the usual 
perturbative series.  We limit ourselves  to the case when fermionic 
dynamics is essentially frozen, considering infinitely massive adjoint quarks 
and studying the static potential between them.  To this purpose we survey
Wilson loops in the adjoint representation. This problem was tackled in 
\cite{paniak} (and at finite temperature in \cite{grignani}, where also many finite volume 
results were derived): there the exact dependence of the string tension 
from the topological parameter labelling the vacuum was found. It was also 
 shown that the same results can be obtained using an effective 
Hamiltonian in the LF theory, in the limit of infinitely massive quarks. To 
perform the computations the authors assumed the point of view, 
suggested by Witten \cite{wittheta}, of simulating topological sectors by defining the 
theory with a Wilson loop at infinity in the $k$-fundamental representation. 
We pursue instead a different approach, 
closer to the instanton interpretation 
of the results and suitable for a  comparison with perturbative physics. 

In Sect.~II we present the general description of  $k$-sectors 
in adjoint $QCD_2$, analysing
the same issue both in the heat-kernel 
(Hamiltonian) language and in the instanton expansion: in the latter case 
the  $k$-dependence arises as a phase factor in summing over inequivalent 
bundle structures in the gauge connection space. In Sect.~III we explicitly
compute  
 the adjoint Wilson loop on the sphere in the $k$-sector for the 
$SU(N)$ theory, as a sum over a set of N integers, 
a representation useful for further developments. We also 
perform the decompactification limit, showing that 
the final result is in full agreement with the one of \cite{paniak}, where the 
computation was done directly on the plane with a Wilson loop at infinity in
the $k$-fundamental representation. 
In Sect.~IV we study
 the  correlation function of an adjoint and a $k$-fundamental
Wilson  loop first on the 
sphere and then in the decompactification limit, in order to mimic the 
inclusion of asymptotic $k$-charges, according to Witten's suggestion.
On the plane we recover the result of the
previous section.
Next we check in Sect.~V that this result is consistent, at the
fourth order 
in the coupling constant, with the LF perturbative computation 
plus the $k$-holonomy at infinity. Sect.~VI  is devoted to derive 
the instanton 
representation on the sphere for both cases, by performing the Poisson 
resummation on 
the exact results: we find that inequivalent gauge bundles enter the game, 
the relative contributions being weighted by the topological parameter.
Nevertheless the instanton patterns turn out to be completely
different
in the two cases. In particular we notice that
the zero-instanton contribution for the former case (adjoint Wilson
loop in the $k$-th sector)
does not depend on the topological number $k$: 
in the decompactification limit it coincides with the sum of the perturbative 
series  
in which the propagator is prescribed according to
Wu-Mandelstam-Leibbrandt (WML) \cite{wu,mandel,leib}, 
for the adjoint Wilson loop 
without the presence of 
the $k$-holonomy at infinity. In the latter case (adjoint Wilson loop
enclosed in a $k$-fundamental one) instead,
string tensions do not depend on $k$ but the 
polynomial part does. 
After checking this fact by using the WML 
propagator in a perturbative expansion at ${\cal O}(g^4)$ in
Sect.~VII, 
we conclude that ${\it only}$ for the complete theory on the 
plane (i.e. full-instanton resummed and then decompactified) the equivalence 
between $k$-sectors and theories with $k$-fundamental Wilson loops at
infinity  holds.
In Sect.~VIII we draw our conclusions and discuss future developments,
whereas technical details are deferred to the Appendix.
 
\section{\kap-sectors and instantons}

It was first noticed by Witten \cite{wittheta} 
that two-dimensional Yang-Mills
theory and two-dimensional QCD with adjoint matter do possess
$k$-sectors. 
We consider $SU(N)$ as the gauge group: since Yang-Mills fields
transform in the adjoint representation, the true local symmetry is
the quotient of $SU(N)$ by its center, $Z_N$. A standard result in
homotopy theory tells us that the quotient is no longer simply
connected, the first homotopy group being $$\Pi_1(SU(N)/Z_N)=Z_N.$$
This result is of particular relevance for the vacuum structure of a
two-dimensional gauge theory: according to the classical picture
\cite{Ja76}, vacuum states are related to static pure gauge configurations
\begin{equation}
A_1=ig(x)^{-1}\partial_1 g(x) ,
\end{equation}
identified when connected by a continuous deformation. 
Assuming boundary conditions that allow for the compactification of the 
space-manifold to $S^1$, we see that the relevant maps $g(x)$ to the gauge 
group $\GG$ fall into equivalence classes labelled by $\Pi_1(\GG)$. 
Starting from 
this observation, standard geometrical and field theoretical arguments
lead to the conclusion that all physical states carry an irreducible
representation of $\Pi_1(\GG)$ \cite{imbo}. In the case at hand we have
exactly  $N$ irreducible 
representations for $Z_N$, labelled by a single integer parameter $k$, 
taking the values $k=0,1,..,N-1$ (it is obviously related to a $N$-th root of 
the identity). On a physical state $|\psi>$ the generator ${\cal C}$ of $Z_N$ 
simply acts  as a phase
\begin{equation}
{\cal C}|\psi>=e^{2\pi i \frac{k}{N}} |\psi>;
\end{equation}
from a gauge theoretical point of view, while physical states are strictly 
invariant under ${\it small}$ gauge transformations (generated by the Gauss' 
law), ${\it large}$ gauge transformations are projectively realized on them. 
Inequivalent quantizations, parametrized by $k$ in the nonabelian (SU(N)) case 
are therefore seen to appear when the matter content singles out the effective 
gauge group, eventually changing the topological properties of the theory 
itself.

We remark that the situation is quite different in the two-dimensional
abelian case: there the homotopy group is $\Pi_1(U(1))=Z$ and
the parameter labelling the irreducible representation is a real
number, taking values between $0$ and $2\pi$: it is usually called
$\theta$ in analogy with the four-dimensional case (we see that the
crucial homotopy group for $QCD_4$ is in fact $\Pi_3(SU(N))=Z$).

Concerning the pure $SU(N)$ Yang-Mills theory, the  explicit solution 
when $k$-states are taken into account was presented in
Ref.~\cite{grignani}:  their 
main result, the heat-kernel propagator on the cylinder, allows to compute 
partition functions and Wilson loops on any two-dimensional compact 
surface, therefore generalizing  the well-known Migdal's 
solution \cite{migdal} to $k$-sectors.

Let us first recall the standard procedure: 
on a two-dimensional cylinder of length $\tau$ and base circle $L$, 
with area $A=L\tau$, one introduces the heat kernel
$${\cal K}[A;{\cal A}_2,{\cal A}_1],$$
${\cal A}_1$ and ${\cal A}_2$ being the potentials at the two boundaries.

It is well-known that, if we introduce the unitary matrices
$$U_{1(2)}={\cal P}\exp\Bigl(i\int_0^{L}dx {\cal A}_{1(2)}(x)\Bigr),$$
${\cal P}$ denoting  path-ordering as usual,
the heat kernel is a class function of $U_1$ and $U_2.$

It enjoys in particular the basic sewing property
\begin{equation}
\label{sewing}
{\cal K}[L\tau;U_2,U_1]=\int dU(u){\cal K}[Lu;U_2,U(u)]
{\cal K}[L(\tau-u);U(u),U_1].
\end{equation}

Thanks to the invariance under area-preserving diffeomorphisms, knowing
${\cal K}[A;U_2,U_1]$, one can easily derive the exact partition function
on the sphere $S^2$ by setting $U_2=U_1={\bf 1}$
$${\cal Z}(A)={\cal K}[A;{\bf 1},{\bf 1}].$$

By expanding ${\cal K}$ in terms of group characters $\chi_R(U)$
\begin{equation}
\label{char}
{\cal K}[A;U_2,U_1]=\sum_R\chi_R(U_1)\chi_R^{\dagger}(U_2)
\exp\Bigl[-\frac{g^2 A}{4}C_2(R)\Bigr],
\end{equation}
R denoting an $SU(N)$ irreducible representation and $C_2(R)$ its quadratic
Casimir, one recovers the well-known expression for the partition function
\begin{equation}
\label{partition}
{\cal Z}(A)=\sum_{R} (d_{R})^2 \exp\left[-{{g^2 A}\over 4}C_2(R)\right],
\end{equation}
$d_R$ being the dimension of the representation $R$ \cite{migdal}. The extension to 
a general compact Riemann surface of genus $G$ without boundaries is trivial, 
consisting in a change of exponent for $d_R$ from 2 to $(2-2G)$.

The generalization of the above construction to $SU(N)/Z_N$ 
is quite simple: following \cite{grignani}, 
we observe that the heat kernel in a 
$k$-sector can be obtained by projecting its  final state 
onto $k$-states. This is done by summing over all transformations of $U_2$ 
by the elements of the center $Z_N$ and weighting each term in the sum
by a phase factor:
\begin{equation}
\label{charna}
{\cal K}_k[A;U_2,U_1]=\sum_{z\in Z_N} z^k \, {\cal K}[A;zU_2,U_1]\,,
\end{equation}
with $z=\exp (2\pi i \frac{n}{N})$, $n=0,\ldots,N-1.$ 
The explicit form of the partition function on the sphere can be
written as follows 
\begin{equation}
\label{partition1}
{\cal Z}_k (A)=\sum_{n=0}^{N-1} e^{2\pi i \frac{nk}{N}}
\sum_R 
\frac{\chi_R\lp e^{-2\pi i \frac{n}{N} } \identity \rp}{d_R} \, 
d_{R}^2 \exp\left[-\frac{g^2
A}4 \, C_2(R)\right]\,.
\end{equation}
It is possible to give a beautiful interpretation of the above expression;
in fact, using the Young tableau representation, one can show that
\beq
\label{series}
\sum_{n=0}^{N-1} e^{2\pi i \frac{nk}{N}}\,
\frac{\chi_R\lp e^{-2\pi i \frac{n}{N} } \identity \rp}{d_R} =
\sum_{n=0}^{N-1} \exp\lq 2\pi i \frac{n}N \lp k -
\sum_{\a=1}^{N-1}m^{(R)}_\a \rp \rq =
\delta_{[N]}(k-m^{(R)})\,,
\eeq
where
\beq
\nome{tab}
m^{(R)}=
\sum_{\a=1}^{N-1}m^{(R)}_\a
\eeq
is the total number of boxes of the Young tableaux and $\delta_{[N]}$
is the $N$-periodic delta function.

The partition function therefore takes the form
\begin{equation}
\label{twist}
{\cal Z}_k(A)=\sum_{R} (d_{R})^2 \exp\left[-\frac{g^2 A}{4}C_2(R)\right]
\delta_{[N]}(k-m^{(R)})\,.
\end{equation}

We now recall that the Casimir $C_2(R)$ is related to the allowed
energies of the system: while in the $SU(N)$ theory all
representations contribute,
in $SU(N)/Z_N$ only a particular class of Casimir invariants appear,
depending on $k$. Different $k$-theories have different energy
spectrum. In particular they exhibit a different ground state, the
$k$-fundamental representation. The modification naturally survives 
the decompactification limit, where it is the $k$-fundamental which
dominates instead of the $1$-fundamental. We remark that the above
characterization of $k$ is, in some sense, nonperturbative, since the
$k$-dependence shows up in  solving  for the Hamiltonian eigenvalues.

On the other hand, it is possible to implement the topological 
dependence even without passing through an energy interpretation.  Let
us consider the familiar case of $QCD_4$ with gauge group
$SU(N)$. There the partition function takes contribution from
four-dimensional connections belonging to disconnected sectors, as the
existence of inequivalent $SU(N)$ bundles over $S^4$
displays (we consider four-dimensional Euclidean space-time
compactified to $S^4$). Inequivalent $SU(N)$ bundles are
classified according to a  single integer, $n$, \ie the Chern number,
related to the field topological charge that is again
determined by $\Pi_3(SU(N))=Z$. Notice that at present we are
discussing genuine four-dimensional vector fields, whereas the
configuration space in the Hamiltonian formalism  
is three-dimensional. If $\ZZ^{(n)}$ is the partition
function in a given sector, we infer for the full answer
\beq
\label{zsector}
\ZZ (\th )=\sum_{n=-\infty}^{\infty} e^{i\omega (n)} \ZZ^{(n)}\,,
\eeq
and, requiring cluster properties, it can be easily shown that
\beq
\label{angle}
\omega(n)=2\pi n \th \,, \quad \quad \th \in [0,2\pi]\,.
\eeq

We can proceed in perfect analogy in two dimensions. To begin with, we realize
that every $SU(N)$ bundle is trivial on $S^2$, since
$\Pi_1(SU(N))=0$. Considering instead $SU(N)/Z_N$, we deduce from
$\Pi_1 \neq 0$ there exists exactly $N$ different, inequivalent
classes of bundles, characterized by $Z_N$ fluxes $n=0,1, \ldots,N-1$.
The presence of the quotient reflects on a richer topological
structure in the connection space, non-trivial bundles entering  the
game.
As in $QCD_4$, we have
\beq
\label{sector2}
\ZZ_k=\sum_{n=0}^{N-1} e^{i\omega(n)} \ZZ^{(n)}\,,
\eeq
$\ZZ^{(n)}$ pertaining to the $n$-th principal bundle. Imposing 
cluster decomposition we get
\beq
\label{angle2}
\omega(n)=\frac{2\pi k}N \,n  \,, \quad \quad  k=0,1,\ldots,N-1\,.
\eeq
We expect perturbative physics to be related to the trivial $n=0$
sector, independent of the topological parameter $k$: as the coupling
$g^2 A \to 0$, all $\ZZ^{(n)}$ except $\ZZ^{(0)}$ should be exponentially
suppressed.
While it is in principle possible to directly compute $\ZZ_k$ by means
of an instanton expansion, namely by finding the classical solution in
any $k$-sector and then expanding the functional integral around it,
we follow here an  alternative route: we obtain a dual version of
Eq.~\re{twist} via a Poisson resummation and there we identify the different
instanton contributions. It turns out Eq.~\re{twist} 
is localized around solutions of Yang-Mills equations of $SU(N)/Z_N$,
as predicted by Witten \cite{witte}: the zero-instanton sector comes
only from the $n=0$ part, reproducing a truly $SU(N)$ perturbative result.

Hence, we have  two complementary representations for the partition
function, in which the parameter $k$ plays different  roles: the
Hamiltonian heat-kernel representation, in which $k$ selects the energies,
and the instanton representation, where $k$ weights the contribution of
inequivalent bundles.

We emphasize that the parallel with the four-dimensional case, or even
with the $U(1)$ model in two dimensions is still not complete: in four
dimensions, in fact, Eq.~\re{zsector} can be reinterpreted as the addition
of the  topological action
\beq
\label{topol}
S_{top}^{(4D)}=\frac{i \th}{16 \pi^2} \int d^4x\, \tr \lq F_{\m\n}
\hat{F}^{\m\n} \rq
\eeq
to the YM term, $\hat{F}^{\m\n}$
being the dual of $F_{\m\n}$, defined in the usual way, and 
 $\frac1{16 \pi^2} \int d^4x\, \tr \lq F_{\m\n}
\hat{F}^{\m\n} \rq=n$ for every connection $A_{\m}$ in the $n$-th sector. 
Analogously, in the two-dimensional case we have to add for $U(1)$
\beq
\label{topol2}
S_{top}^{(2D)}=\frac{i\th}{4\pi} \int d^2x\,  F_{\m\n}\,
\e^{\m\n}\,.
\eeq
These modifications are motivated by an  instanton expansion
and are general, insensitive to whether the theory is defined on a
compact  surface
or in the Euclidean space. 
Nevertheless, inspired by the $U(1)$ case, Witten suggested a way to
implement the phase factor (or equivalently the $k$-dependence)  
in Eq.~\re{sector2}  directly on the plane, using the $SU(N)$ theory. The
argument goes as follows: as pointed out long ago by Coleman \cite{cole},
in the Schwinger model one introduces the parameter $\th$ as
the strength of a fractional charge $\frac{e\th}{2\pi}$ at the
spatial right end of the two-dimensional world and its opposite at the
left end. 
We observe that the inclusion of the topological term  can be understood as
imposing the following  generalized boundary condition: we can interpret such
charges in terms of a Wilson loop enclosing the world, which can be
explicitly included in the action of the theory
\beeq
\label{extaction}
\ZZ_{\th}&=&\int \DD \bar{\psi} \DD \psi \DD A_{\m} \, \exp \lq - \int d^2x\,
{\cal L} + \frac{i \th}{2\pi} \int_{C_\infty} dx^{\m}A_{\m} \rq \no \\
&=&\int \DD \bar{\psi} \DD \psi \DD A_{\m} \, \exp \lq - \int d^2x\,
{\cal L} + \frac{i\th}{4\pi}\int  d^2x \, F_{\m\n} \e^{\m\n}\rq \,.
\eeeq
In the nonabelian case the suggestion is to consider static colour
charges $T_R$ and $T_{\bar{R}}$ at the boundary. Here the $T$'s are
the generator of the nonabelian colour group in the representation $R$
and its conjugate, respectively. Unlike the abelian case, there is no
continuous parameter, the only choice is discrete, depending on the
representation of the boundary charges. If $\tr_R$ is the trace in the
representation $R$ of the gauge group, it holds
\beq
\label{charges}
\ZZ_{(R)}=\int \DD \bar{\psi} \DD \psi \DD A_{\m} \, \exp \lq - \int d^2x\,
{\cal L} \rq \tr_R \,{\cal P} \exp \lp 
i \int_{C \to \infty} dx^{\m}A_{\m} \rp\,,
\eeq
which manifests that
the Wilson loop at infinity cannot be written as a local addition to
the Lagrangian. Let us notice that the theory was written in presence
of adjoint dynamical fermions. Witten showed that this system is
stable (there are no energetically favoured pair creations in the external
background) only when $R$ is one of the $N$ (antisymmetric)
fundamental representation of $SU(N)$, that are exactly the $k$-ground
states appearing in Eq.~\re{sector2}. We remark that $\ZZ_{(R)}$ is
computed with $A_{\m}$ taking values in the $SU(N)$ algebra,
and  has to be used to evaluate general observables (like a Wilson loop
in the adjoint representation) in the absence of dynamical fermions.

In the next section we explicitly check
that this prescription is correct and that Wilson loops computed on
$S^2$ in the $SU(N)/Z_N$ theory match, in the decompactification
limit, with the result of Ref.~\cite{paniak}, where the above picture was
adopted. 
However, when remaining on the sphere, \ie before taking such a limit,
we do not see any reason why such a procedure should lead to the
correct  $SU(N)/Z_N$
result. A $k$-fundamental loop on $S^2$ for $SU(N)$ cannot mimic
non-trivial  bundles, as it cannot obviously modify its instanton
structure.
Thus we do not expect that the zero-instanton result for an adjoint
loop in this context will  coincide with the genuine zero-instanton
term for $SU(N)/Z_N$, neither before nor after decompactification.

\section{The Wilson loop in the adjoint representation}

\noindent

Starting from the sewing property of the heat kernel Eq.~(\ref{sewing}),
it is also easy
to get an expression for a Wilson loop winding around a smooth non
self-intersecting closed contour on $S^2$.  By choosing again $U_1=U_2=
{\bf 1}$ and by inserting the Wilson loop expression for a contour
in a given representation $T$, we get
\begin{eqnarray}
\label{wilson}
{\cal W}(A_1,A_2)&=&\frac1{{\cal Z}(A)d_T} \sum_{R,S} d_{R}d_{S}
\exp\left[-\frac{g^2 A_1}{4}C_2(R)-\frac{g^2 A_2}{4}C_2(S)\right]
\no \\
&\times & \int dU \, {\rm Tr}_{T}[U]
\, \chi_{R}(U) \chi_{S}^{\dagger}(U),
\end{eqnarray}
$A_1$ and $A_2$ ($A_1+A_2=A$) being the areas singled out by the loop.

A particularly interesting case is represented by the choice of the
loop in the adjoint representation $T\equiv adj$; in so doing
invariance under the quotient group is preserved. On the other hand the 
loop might somehow mimic contributions from the would-be
``transverse'' vector degrees of freedom in higher dimensions.

The projection on a given sector $k$ can now again be realized by
``twisting'' $U_2={\bf 1}$ with the center factor $z_k$
\beeq
\label{wilsonk}
\wk (A_1,A_2)&=&\frac1{\zk \, (N^2-1)} \sum_{R,S} d_{R}d_{S}
\exp\left[-\frac{g^2 A_1}{4}C_2(R)-\frac{g^2 A_2}{4}C_2(S)\right]
\no \\
&\times & \int dU \, {\rm Tr}_{adj}[U]
\, \chi_{R}(U) \chi_{S}^{\dagger}(U)\,
\de_{[N]} \lp k- m^{(S)} \rp\, .
\eeeq

In the decompactification limit $A\to \infty$, keeping $A_1$ fixed,
the above quantity is to be interpreted as the Wilson loop average
in a $k$-vacuum, for the theory defined on the plane.

Our next step will consist in
working Eqs.~(\ref{twist},\ref{wilsonk}) out to cast them in a
desirable form. In addition, we anticipate in passing that the
instanton representation, which will be the subject of Sect.~VI,
hinges precisely on the formulae we will eventually arrive at.
Taking the well-known relation 
\beq 
\nome{adj}
{\rm Tr}_{adj}[U]={\vert \tr U\vert}^2-1\,,
\eeq
into account,
Eq.~\re{wilsonk} becomes
\beeq
\label{wilson1}
\frac1{N^2-1}+ \wk (A_1,A_2)&=&\frac1{\zk \, (N^2-1)} \sum_{R,S} d_{R}d_{S}
\exp\left[-\frac{g^2 A_1}{4}C_2(R)-\frac{g^2 A_2}{4}C_2(S)\right]\\
&\times & \int_0^{2\pi} 
d\th_1 \ldots d \th_N \, \de \lp \sum_{j=1}^N \th_j\rp\,
\sum_{p,q=1}^{N} e^{i (\th_p-\th_q)}\;
\de_{[N]} \lp k- m^{(S)} \rp\,. \no
\eeeq
We now proceed as follows. Firstly, we switch to the integers
$\hat{l}_q$ so defined
\beq
\nome{lindex}
\hat{l}_q=m_q-q+N\,, \qquad \quad q=1, \ldots, N-1\,,
\eeq
which satisfy the $SU(N)$ constraint $\hat{l}_1>\hat{l}_2> \ldots 
\hat{l}_{N-1}>0$, turning a weakly monotonous sequence into a strongly
monotonous one.
Secondly, with the twofold purpose of  extending the range of the
$\hat{l}_q$'s, $ q=1, \ldots, N-1\,,$ also to negative integers and
of gaining  the symmetry over permutations of a full set of $N$
indices in Eqs.~(\ref{twist},\ref{wilson1}),
we introduce the obvious equality ($\hat l_N$ is here a dummy quantity)
\beq
\nome{equality}
\sqrt \pi= \int_0^{2\pi}d\a\, \sum_{\hat{l}_N=-\infty}^{+\infty}
e^{-\lp \a -\frac{2\pi}N \sum_{j=1}^{N-1} \hat{l}_j -2\pi
\hat{l}_N\rp^2}\,. 
\eeq
Thanks to it, we extend the set of representation indices by defining
\beeq
\nome{defl} 
l^{R,S}_q &=& \hat{l}^{R,S}_q  + \hat{l}^{R,S}_N \,,
\qquad \quad  q=1, \ldots, N-1\,,\no \\
l^{R,S}_N &=& \hat{l}^{R,S}_N\,,\no \\
l^{R,S}   &=& \sum_{i=1}^N l^{R,S}_i \,.
\eeeq
The operations hitherto carried out, enable us to write
Eqs.~(\ref{twist},\ref{wilson1}) explicitly in terms of the new set of
indices $l_i=(l_1, \ldots, l_N)$. By recalling the relations
\begin{eqnarray}
\label{casimiri}
C_2(R)&=&
\sum_{i=1}^{N} \lp l_{i}-\frac{l}{N} \rp^2-
\frac{N}{12}(N^2-1)
\nonumber \\
d_{R}&=&\Delta(l_1,...,l_{N})\,,
\end{eqnarray}
where $\Delta$ is the Vandermonde determinant , we get
\beeq
\label{partip}
&&\zk (A)=\frac{(2\pi)^{N-1}}{N!\,\sqrt\pi } \sum_{l_i=-\infty}^{+\infty} 
\int_0^{2 \pi}
d\a\, e^{-\lp \a - \frac{2\pi}N l \rp^2}
\de_{[N]} \lp k- l + \frac{N(N-1)}2 \rp \no \\
&& \ph{\zk (A)=\frac1{N!\,\sqrt\pi } \sum_{l_i=-\infty}^{+\infty} } \quad
\times
\exp \left [ -\frac{g^2 A}{4} C_2 (l_i)\right ]
\Delta^2(l_1,...,l_N)
\eeeq
and
\beeq
\label{wilsonp}
&&\frac1{N^2-1}+ \wk (A_1,A_2)=\frac1{\zk \, (N^2-1)}
\sum_{l_i^R,\, l_i^S=-\infty}^{+\infty} \frac1{\pi (N!)^2}
\int_0^{2\pi} d\th_1 \ldots d \th_N \, \de \lp \sum_{j=1}^N \th_j\rp\no \\
&&\times
\int_0^{2 \pi}
d\a_1 \,d\a_2 \, \,e^{ -\lp \a_1 - \frac{2\pi}N l^R \rp^2}\,
e^{-\lp \a_2 - \frac{2\pi}N l^S \rp^2 }
\exp\left[-\frac{g^2 A_1}{4}C_2(l^R_i)-\frac{g^2 A_2}{4}C_2(l^S_i)\right] \\
&&\times \sum_{p,q=1}^{N} e^{i (\th_p-\th_q)}\;
\prod_{h=1}^N e^{i l_h^R \th_h} \prod_{r=1}^N e^{-i l_r^S \th_r} \,
\de_{[N]} \lp k- l^S + \frac{N(N-1)}2 \rp 
\Delta(l^R_1,...,l^R_N) \Delta(l^S_1,...,l^S_N) \no\,.
\eeeq

It is convenient to interpret the constraint on the angles $\theta_i$
in the equation above, as 
a periodic $\de$-distribution
\beq
\nome{delta}
\de \lp \sum_{j=1}^N \th_j\rp = \frac1{2\pi} \sum_{n=-\infty}^{+\infty}
e^{in\sum_{j=1}^N \th_j}\,\,,
\eeq
the total volume being still finite and occurring 
also in the
partition function at the denominator. 

It is now natural to perform the following shift in Eq.~\re{wilsonp}
\beq
\nome{shift}
l_j^R \rightarrow l_j^R +n \,,
\eeq
under which both $\Delta (l_j^R)$ and $C_2(l^R_i)$ are
insensitive.

Hence, Eq.~\re{wilsonp} reads
\beeq
\label{wilsonfin}
&&\frac1{N^2-1}+ \wk (A_1,A_2)
=\frac1{\zk \, (N^2-1)}
\sum_{l_i^R,\, l_i^S=-\infty}^{+\infty} 
\frac1{2\pi^2 (N!)^2}
\sum_{n=-\infty}^{+\infty}
\int_0^{2\pi} d\th_1 \ldots d \th_N \, \no \\
&&\times
\int_0^{2 \pi}
d\a_1 \,d\a_2 \, \, e^{ -\lp \a_1 - \frac{2\pi}N l^R + 2\pi n\rp^2}\,
e^{-\lp \a_2 - \frac{2\pi}N l^S \rp^2 }
\exp\left[-\frac{g^2 A_1}{4}C_2(l^R_i)-\frac{g^2 A_2}{4}C_2(l^S_i)\right] \\
&&\times \sum_{p,q=1}^{N} e^{i (\th_p-\th_q)}\;
\prod_{h=1}^N e^{i l_h^R \th_h} \prod_{r=1}^N e^{-i l_r^S \th_r} \,
\de_{[N]} \lp k- l^S + \frac{N(N-1)}2 \rp 
\Delta(l^R_1,...,l^R_N) \Delta(l^S_1,...,l^S_N) \no\,.
\eeeq
The next advance in the computation of $\wk (A_1,A_2)$ is  reached by
implementing Eq.~\re{equality} backwards and by 
working out the integration over the $N$ angles $\th_i$, which produces
\beq
\nome{intheta}
(2 \pi)^N  \, N! \,\sum_{q_1,q_2=1}^{N} \, \prod_{j=1}^N \de (l_j^R - l^S_j 
+\de_{j,\, q_1}-  \de_{j,\, q_2} )\,,
\eeq
implying, for $q_1$, $q_2$ fixed 
\beq
\nome{lr}
l^R_j=l^S_j-\de_{j,\, q_1}+\de_{j,\, q_2}\,.
\eeq
Although harmless as far as $l^R$ is concerned, such a shift
affects both the Casimir and the Vandermonde determinant related to
the $R$ representation. In particular, for the former we have
\beq
\nome{shiftcasi}
\begin{array}{ll}
C_2(l^R)=C_2 (l^S) + 2 \, (l^S_{q_2} - l^S_{q_1} +1 ) & \quad {\rm
if} \; \; \; q_1\neq q_2 \\
C_2(l^R)=C_2 (l^S) & \quad {\rm if} \;\;\; q_1 = q_2 \,,
\end{array}
\eeq
leading to  the following form for $\wk$
\beeq
\label{wilsonint}
&&\wk (A_1,A_2 )=\frac1{N+1}
\Biggl\{ 1+ 
\frac2{\zk \, (N-1)}\,
\frac{(2\pi)^{N-1}}{\sqrt \pi \,N!}
\sum_{l_i=-\infty}^{+\infty} \,
\sum_{1=q_1<q_2}^{N} \exp \lq - \frac{g^2 A_1}2 \lp l_{q_2}
-l_{q_1}+1 \rp \rq 
\no \\
&&\ph{\wk (A_1,A_2 )=\frac1{N+1} } 
\times 
\int_0^{2 \pi}
d\a \,\,
e^{-\lp \a - \frac{2\pi}N l  \rp^2} 
\de_{[N]} \lp k- l + \frac{N(N-1)}2 \rp  \no \\
&&\ph{\wk (A_1,A_2 )=\frac1{N+1} }\times 
\Delta(l_1,...,l_N) \, \Delta(l_1,\ldots ,  l_{q_1} -1,
\ldots,   l_{q_2}+1 , \ldots, l_N)  
\Biggr\}  
\,,
\eeeq
where also the normalization to $\zk$ of Eq.~\re{partip} has been explicitly
carried out in the first term of the summation.
At this stage we possess nice formulae which enable us to deal with
the decompactification limit.

The partition function \re{twist} and the Wilson loop \re{wilsonk} 
in the limit of infinite area of the sphere are dominated by
particular representations labelled by suitable indices $\{\bar l_i\}$.
It is now easy to see that the dominant contribution is given by the
following set 
\beq \nome{dom}
\{\bar{l}_i\}=\lgr 0,\, 1, \, 2, \, \ldots,\, N-k-1, \,N-k+1,\,\ldots,
N-1,\, N \rgr\,,
\eeq
and their permutations, which obey $\bar l= k+ \frac{N(N-1)}2$ and for
which  the minimum value of the Casimir is reached 
\beq\nome{mincas}
C_2(\bar{l}_i)=\frac{k(N-k)(N+1)}N\,.
\eeq
The number  of distinct permutations, leaving the Casimir Eq.~\re{mincas}
unchanged, amounts to ${N \choose k}$.
Evaluating $\zk$ for $l_i=\bar{l}_i$, we obtain 
\beeq\nome{partdec} 
\zk (A\to \infty ) 
&=&\frac{(2\pi)^{N-1}}{N!\,\sqrt\pi } \,
\Delta^2(\bar{l}_1,...,\bar{l}_N)
\int_0^{2 \pi}
d\a\, \exp \lq -\lp \a - \frac{2\pi}N  k - \pi (N-1) \rp ^2 \rq \no \\
&\times &
{N \choose k}
\exp \lq -\frac{g^2A}4  \, \frac{k(N-k)(N+1)}N \rq \,.
\eeeq
We observe that in the last line of Eq.~\re{partdec} the exponent
coincides with  the Casimir of the $k$-fundamental
representation, multiplied by $g^2A/2$, and ${N \choose k}$ with its dimension.

We now focus on  the Wilson loop in the form of
Eq.~\re{wilsonint}. The decompactification limit $A \to \infty$, $A_1$
fixed, is performed by evaluating $\wk$ on the configurations
\re{dom}. A simple form is found when the dependence on $\zk$ is
factorized  out via Eq.~\re{partdec}
\beeq
\nome{wkdecren}
\wk (A_1,A_2 \to \infty)=\frac1{N+1} \Biggl\{ 
1 &+& \frac1{N-1}\,
\sum_{q_1\neq q_2=1}^N
\frac{\Delta(\bar l_1,\ldots , \bar l_{q_1} -1,
\ldots,  \bar l_{q_2}+1 , \ldots, \bar l_N)}{\Delta(\bar l_1,...,\bar
l_N)} \no \\ 
& \times & 
\exp \lq - \frac{g^2 A_1}2 \lp \bar{l}_{q_2}
-\bar{l}_{q_1}+1 \rp \rq  \Biggr\}\,.
\eeeq
This is not yet the end of the story.
At this stage we  have still to specify  what are the shifts in $\bar
l_j$ allowed by
the  Vandermonde determinants in the 
last term of Eq.~\re{wkdecren} and determine the string tensions
$\s (q_1,q_2)$, to be read from the exponential in the same
equation~\footnote{We define the string tension
$\s$ to be the exponent with changed sign divided by half the area of the
loop.},  and the relative weights  they give rise to.
It turns out the following four cases can occur
\begin{itemize}
\item
$q_1=j$ and $q_2=j-1$, with
$j=1,\,\ldots,\,N-k-1,\,N-k+2,\,\ldots,\,N$. As a whole, there are
$N-2$ possible swaps with vanishing string tension. Each of them
contributes with the same weight;
we can  for instance choose $q_1=2$ and $q_2=1$ 
$$
\frac{\Delta(\bar l_2, \bar l_1,\ldots,\bar l_N)}
{\Delta(\bar l_1,\bar l_2, \ldots,\bar l_N)} = -1
$$
\item 
$q_1=1$ and $q_2=k$, with string tension $\s (1,k)=\frac{g^2}2 (N-k)$ and 
weight
$$
\frac{\Delta(\bar l_1-1,\ldots,\bar l_k +1 , \bar l_{k+1}\ldots,\bar l_N)}
{\Delta(\bar l_1,\bar l_2, \ldots,\bar l_N)} =
\frac{(N+1)(N-k-1)}{k+1}
$$
\item 
$q_1=k+1$ and $q_2=N$, with string tension $\s (k+1,N)=\frac{g^2}2 k$
and weight
$$
\frac{\Delta(\bar l_1,\ldots,\bar l_k, \bar l_{k+1} -1,\ldots,\bar l_N+1)}
{\Delta(\bar l_1,\bar l_2, \ldots,\bar l_N)} =
\frac{(N+1)(k-1)}{N-k+1}
$$
\item
$q_1=1$ and $q_2=N$, with string tension $\s (1,N)=\frac{g^2}2 (N+1)$
and weight
$$
\frac{\Delta(\bar l_1-1,\bar l_2, \ldots,\bar l_N+1)}
{\Delta(\bar l_1,\bar l_2, \ldots,\bar l_N)} =
\frac{k N (N+2)(N-k)}{(k+1)(N-k+1)}\,.
$$
\end{itemize}
Finally, by  substituting the previous results in Eq.~\re{wkdecren}, 
$\wk (A_1,A_2 \to \infty)$ becomes
\beeq
\nome{wkdecfin}
\wk (A_1,A_2 \to \infty)&=&\frac1{N^2-1} \lq 1+
\frac{k N (N+2)(N-k)}{(k+1)(N-k+1)}
e^{-\frac{g^2 A_1}2 \,(N+1)}
\right. \\
&+& \left.
\frac{(N+1)(N-k-1)}{k+1}
e^{-\frac{g^2 A_1}2 \,(N-k)}+ 
\frac{(N+1)(k-1)}{N-k+1}
e^{-\frac{g^2 A_1}2 \,k}
\rq\,. \no
\eeeq
A comment is now in order. Our result Eq.~\re{wkdecfin} coincides with
Eq.~(14) of Ref.~\cite{paniak}, which was derived following an
alternative route. In fact, each term in the sum \re{wkdecfin}
corresponds to an irreducible $SU(N)$ representation into which  
the tensor product of an adjoint representation with a
$k$-fundamental one is decomposed. To realize this, notice that 
the overall normalization is but the
dimension of the adjoint, the prefactor of each term denotes the degeneracy
(dimension of the representation normalized to ${N \choose k}$) 
and the exponent is the Casimir
multiplied by $g^2 A_1/2$.
Nevertheless, we emphasize that such a procedure is applicable only
in the decompactification limit. As opposed to this, our starting point,
Eq.~\re{wilsonint}, holds in  more general instances and will be
unavoidable in considering the instanton representation.

\section{Two-loop correlation on the sphere}

\noindent
As promised, in this section we consider the correlation on the sphere
between two non-intersecting (nested) loops, one in the adjoint representation
and the other in the $k$-fundamental one. 
We call $A_2$ the area 
of the annulus between the loops, $A_3$ and $A_1$ the other two areas
encircled by the loops so that $A_1+A_2+A_3$ equals the total area $A$
of the sphere. We shall firstly take
the decompactification
limit $A_3\to \infty$, keeping fixed the other two areas. Eventually 
we shall 
send the $k$-fundamental 
loop to $\infty$, by performing the
second limit $A_2\to \infty$ keeping $A_1$ fixed.

Our purpose in so doing is to explore to what extent this procedure
reproduces the result we have obtained in Sect.~III, working with a
single loop in the adjoint representation on the sphere in a $k$-sector.
As anticipated in the Introduction, we find
that the Witten conjecture, namely, in our language, that the two results
have to coincide when the sphere is decompactified to the plane, is indeed
verified. The $k$-selection rule on the allowed representations
can be interpreted as the presence of $k$-fundamental
charges at $\infty$. However this is true only for the exact solution,
and only after decompactification of the sphere.

Following the equations given for the heat kernel in Sect.~II, 
it is easy to write the expression for the correlation of the two loops 
on the sphere we mentioned (we use the short-hand notation $C_R\equiv
C_2(R)$)
\beeq
\label{corr}
&&{\cal W}_{k,a}(A_1,A_2,A_3)=\frac{1}{{\cal Z}_k
{\cal W}_k\,(N^2-1)}\sum_{R,S,T}
\exp\Big[-\frac{g^2}{4}(A_3C_R+A_2C_S+A_1C_T)\Big]\no \\
&&d_Rd_T\,\int\,dU_1\, \chi^\dagger_R(U_1)\, \chi_k(U_1)\,\chi_S(U_1)\int\,dU_2
\,\chi^\dagger_S(U_2)\,\chi_a(U_2)\,\chi_T(U_2),
\eeeq
where
\beq
\label{nor}
{\cal Z}_k\,{\cal W}_{k}=\sum_{R,S}d_Rd_S
\exp\Big[-\frac{g^2}{4}(A_3C_R+(A-A_3)C_S)\Big]
\int\,dU\,\chi^\dagger_R(U)\,\chi_k(U)\,\chi_S(U),
\eeq
so that the natural normalizations $${\cal W}_{k,a}(A_1=0,A_2,A_3)=1\,,$$
$${\cal W}_k(A_3,A-A_3=0)=1$$
will ensue. We notice we have dropped the common factor $N\choose k$;
further irrelevant common factors will be 
dropped in the following.

To warm up we  begin considering ${\cal W}_k$
\beeq
\label{wilsonknor}
&&\wk (A_3,A-A_3)=\frac{k!}{{\cal Z}_k (A)}\,\sum_{R,S} d_{R}d_{S}
\exp\left[-\frac{g^2 A_3}{4}C_R-\frac{g^2 (A-A_3)}{4}C_S\right]\\
&&\ph{\wk (A_3,A-A_3)=}
\int_0^{2\pi} 
d\th_1 \ldots d \th_N \, \de \lp \sum_{j=1}^N \th_j\rp 
\sum_{j_1<\ldots<j_k} e^{i (\th_{j_1}+\ldots+\th_{j_k})}\,
{\rm det}||e^{i\hat l^R_h\th_p}||\,{\rm det}||e^{-i\hat l^S_r\th_q}||\,.\no
\eeeq
We now repeat the familiar procedure, integrating over the angles and 
taking symmetry properties into account; we end up with the expression
\beeq
\label{wilsonkhar}
&&\wk (A_3,A-A_3){\cal Z}_k(A)=\sum_{l_q=-\infty}^{+\infty}
\int_{-\infty}^{+\infty}dl\,d\beta
\exp\lq i\beta \lp l-\sum_{j=1}^N l_j \rp \rq \times \no \\
&&
\int_{0}^{2\pi}d\alpha\,
\exp\lq -\lp\alpha- \frac{2\pi}{N}l \rp^2 \rq
\exp\left[-\frac{g^2 A_3}{4}C(l_j^R)-\frac{g^2 (A-A_3)}{4}C(l_j)\right]
\Delta(l_j^R)\,\Delta(l_j),
\eeeq
where
\beq
\label{vinc}
l_1^R=l_1-1,\,\ldots \,, \, l_k^R=l_k-1, \, l_{k+1}^R=l_{k+1},\, 
\ldots\,, \, l_N^R=l_N
\eeq
and $C(l_j)$ is the usual $SU(N)$ expression of the quadratic Casimir
in terms of the representation labels. 

Next, we go back to Eq.~(\ref{corr}); performing the usual harmonic analysis
in terms of Young tableaux and taking symmetry properties into account,
it can be written as
\beeq
\label{corre}
&&{\cal W}_{k,a}(A_1,A_2,A_3)=\frac{1}{N+1}+\frac1{{\cal Z}_k
{\cal W}_k\,(N^2-1)}{N\choose k}\sum_{l_j=-\infty}^{+\infty}
\sum_{q_1\ne q_2 =1}^N
\int_{0}^{2\pi}d\alpha \exp \Big[-(\alpha-\frac{2\pi}{N}l)^2\Big] \no\\
&&\times \exp\Big[-\frac{g^2}{4}(A_3C(l_j^R)+A_2C(l_j)+A_1C(l_j^T))\Big]
\Delta(l_j^R)\,\Delta(l_j^T)
\eeeq
with the constraints $$l_1^R=l_1-1,\,\ldots,\,l_k^R=l_k-1,\,l_{k+1}^R=l_{k+1},
\,\ldots,\,l_N^R=l_N,$$ $$l_1^T=l_{q_1}-1,\,l_2^T=l_{q_2}+1,\,l_j^T=l_{q_j}$$
for $j=3,\ldots,N$ and $$l=\sum_{j=1}^N l_j\,.$$
Eqs.~\re{wilsonkhar} and \re{corre} are suitable for considering the
decompactification limit $A_3\to \infty$, $A-A_3$ fixed and 
$A_3\to \infty$, $A_2$, $A_1$ fixed, respectively. 

Let us start from Eq.~\re{wilsonkhar}.
Since the constraints \re{vinc} imply
\beq\label{cas2}
C_2(l^R)=\sum_{i=1}^k \lp l_i-1-\frac{l}N \rp^2+ \sum_{i=k+1}^N
\lp l_i-\frac{l}N\rp^2-\frac{k^2}N -\frac{N(N^2-1)}2\,,
\eeq
it is immediately recognized that $W_kZ_k$ is dominated by particular
representations labelled by the following set of indices
\beq \nome{dom2}
\{\bar{l}_i\}=\lgr 0,\, 1, \, 2, \, \ldots,\, N-k-1, \,N-k+1,\,\ldots,
N-1,\, N \rgr\,,
\eeq
and their permutations, for which  the minimum value of the
$R$-representation Casimir is reached ($C_2(\bar{l}^R)=0$).
Correspondingly we have
\beq\label{cas22}
C_2(\bar{l})=\sum_{i=1}^N \lp \bar{l}_i-\frac{\bar{l}}{N}\rp^2- 
\frac{N(N^2-1)}2= \frac{k(N-k)(N+1)}N\,,
\eeq
\ie the Casimir of the $k$-fundamental representation.   
Notice that the dominant  configurations of indices coincide with the
ones we found in Sect.~III, when performing the decompactification limit
of the Wilson loop $\wk (A_1,A_2)$ in the $k$-th sector.

Proceeding further with ${\cal W}_{k,a}(A_1,A_2,A_3)$, the evaluation
of Eq.~\re{corre} for $l_i=\bar{l}_i$ yields
\beeq\label{corredom}
{\cal W}_{k,a}(A_1,A_2,A_3\to \infty)&=&\frac{1}{N+1}+\frac1{(N^2-1)}
\sum_{q_1\neq q_2=1}^N
\frac{\Delta(\bar l_1,\ldots , \bar l_{q_1} -1,
\ldots,  \bar l_{q_2}+1 , \ldots, \bar l_N)}{\Delta(\bar l_1,...,\bar
l_N)} \no \\
\phantom{{\cal W}_{k,a}(A_3\to \infty,A_2,A_1)}
&\times &
\exp \lq - \frac{g^2 A_1}2 \lp \bar{l}_{q_2}
-\bar{l}_{q_1}+1 \rp \rq  \,,
\eeeq
having taken the constraints on $\bar{l}_j^T$ into account.
This equation exhibits the remarkable property of being independent of
$A_2$. Indeed, it precisely coincides with Eq.~\re{wkdecren} of Sect.~III, 
which
leads straight to Eq.~\re{wkdecfin}. The independence of $A_2$ of
Eq.~\re{corredom} justifies the procedure adopted in Ref.~\cite{paniak},
where string tensions and weights were obtained just by decomposing
the direct product of an adjoint and a $k$-fundamental representation
into its irreducible components.

At this point, we would like to stress some remarkable features of
Eq.~\re{wkdecfin}. If the  adjoint loop 
is interpreted as a pair of adjoint fermions in the $k$-th vacuum state,
we immediately see that the addition of the $k$-loop at infinity gives rise 
to up to four distinct
singlet configurations: the interaction energies between adjoint
charges generally depend on the vacuum $k$ in which they are measured.
Furthermore,
the physics of the adjoint loops depends on $k$
{\it mod} $N$, in analogy with the continuous $\th$-angle of the Schwinger
model, which is periodic in $2\pi$, and
is symmetric under changing the
representation of the external loop by $k \to N-k$ (which in turn goes
back to the
invariance of the adjoint representation under charge conjugation).
It follows that the vacua corresponding to $k$ and $N-k$ are degenerate in
energy.
Finally, 
surprisingly enough, a configuration presents vanishing
string  tension and  hence the binding
energy between the static adjoint charges is vanishing in the
non-trivial  $k\neq 0$ sectors of the theory.

We have thus proven Witten's conjecture, namely that 
the $SU(N)/Z_N$ theory on the plane in a $k$-sector is equivalent to the 
usual $SU(N)$ theory in presence of a $k$-fundamental Wilson loop at 
infinity.

\section{Perturbative series defined via CPV prescription}

\noindent
We now check that the expectation value of an adjoint loop enclosed in
an asymptotic  $k$-fundamental one  on the plane, expressed by
Eq.~\re{wkdecfin},  is consistent with the
perturbative LF computation, at least up to $O(g^4)$. Such a calculation
suffices to give a flavour on how things work at higher order.

To begin with, let us briefly review the outset of quantization in the
{\em light-cone} gauge $A_{-}=0$  in two dimensions.
If the theory is quantized on the {\em light-front} (at equal $x^{+}$),
no dynamical degrees of freedom occur as the non-vanishing 
component of the vector field does not propagate
\begin{equation}
\label{CPVprop}
D^{P}_{++}(x)=-\frac{i}{4}|x^{-}|\,\delta(x^{+}),
\qquad\qquad x^{\pm}=\frac{x^0\pm x^1}{\sqrt2}\,,
\end{equation}
but rather gives rise to an instantaneous (in $x^{+}$) Coulomb-like
potential. A formulation based essentially on the potential in Eq.~\re{CPVprop}
was originally proposed by G. 't Hooft in 1974 \cite{hooft}, to derive
beautiful solutions for the $q\bar q$-bound state problem under the form of
rising Regge trajectories.

On the other hand, when the theory is quantized at {\em equal-times}, the  
free propagator has the following
causal expression (WML prescription) in two dimensions
\begin{equation}
\label{WMLprop}
D^{WML}_{++}(x)={1\over {4\pi}}\,\frac{x^{-}}{-x^{+}+i\epsilon x^{-}}\,,
\end{equation}
first proposed by T.T. Wu \cite{wu}. In turn this propagator is nothing
but the restriction in two dimensions of the expression proposed 
by S. Mandelstam
\cite{mandel} and G. Leibbrandt \cite{leib} in four
dimensions~\footnote{In dimensions higher than two, where physical
degrees of freedom are switched on (transverse ``gluons''), this
causal prescription is the only acceptable one \cite{libro}.}  and 
derived by means of a canonical quantization in Ref.~\cite{bosco}.

When inserted in perturbative Wilson loop calculations, expressions 
\re{CPVprop} and \re{WMLprop} lead to completely different
results, as first noticed in Ref.~\cite{noi1}. The origin of this 
discrepancy was eventually clarified 
in Ref.~\cite{capo}, where it
was shown that genuine non-perturbative excitations (``instantons'')
are necessary in the ET formulation (Eq.~\re{WMLprop})
in order to obtain the  exact result, which in turn is easily
recovered in the LF formulation (Eq.~\re{CPVprop})
just by summing the perturbative series. 

As announced, we presently concentrate on CPV prescription Eq.~\re{CPVprop},
whereas the WML prescription will be discussed in Sect.~VII.
Firstly, let us start from the perturbative definition  of $\wka$ in
the light-cone gauge
\beeq\nome{wcurr}
&&\wka=\frac1{(N^2-1)\,\zk\wk} \,
\Bigg\{ \tr {\cal P} \exp \lq g \oint_{\G_k} t^b \frac{\de}{\de\, J^b(x)} 
\, \, dx^{+}\rq 
\tr {\cal P} \exp \lq g \oint_{\G_a} T^a \frac{\de}{\de\, J^a(x)} \, \, 
dx^{+}\rq \no \\
&&\ph{\wk=\frac1{(N^2-1)\,\zk}  
\Bigg\{ \tr }\times
\exp \lq -\half \int d^2x \, d^2y \,   J^c(x) \, D(x-y)\,
J^c(y) \rq  \Bigg\}_{J=0} \,,
\eeeq
where  the propagator $D(x-y)$ is defined through Eq.~\re{CPVprop} 
and the matrices $T^a$, $t^b$ belong to the adjoint and to the
$k$-fundamental representations, respectively. Notice that
normalization is such that when $\G_a$ is shrunk to a point, $\wka=1$.
Next we consider two light-like rectangles (see Fig.~1), one with
sides $2l$, $2t$, where the adjoint representation sits, 
nested in a larger rectangle with sides $2L$, $2T$, where instead the
$k$-fundamental sits, 
and choose the currents with support on the contours, so that
$$
J^a(x^+,x^-)= j_1^a \, \de(x^- - L) + j_2^a \, \de(x^- + L)\,
+ j_3^a \, \de(x^- - l) + j_4^a \, \de(x^- + l)\,.
$$
With this choice the perturbative expansion Eq.~\re{wcurr} for $\wka$
reads
\beeq\nome{variexp}
\wka&=& \frac1{(N^2-1)\,\zk\wk} \,  \tr \Biggl\{
{\cal P} \exp \lq g \oint_{C_{\,2}} t^{b_2} 
\frac{\de}{\de\, j_2^{b_2}(x^+)} \, dx^{+}\rq 
{\cal P} \exp \lq g \oint_{C_{\,1}} t^{b_1} \frac{\de}{\de\,
j_1^{b_1}(x^+)}\, dx^{+}\rq  \Biggr\} \no \\
&\times &\tr \Biggl\{ {\cal P} \exp \lq g 
\oint_{C_{\,4}} T^{a_2} 
\frac{\de}{\de\, j_4^{a_2}(x^+) }\, dx^{+}\rq 
{\cal P} \exp \lq g \oint_{C_{\,3}} T^{a_1} 
\frac{\de}{\de\, j_3^{a_1}(x^+)}\, dx^{+}\rq \Biggr\}  \no \\
&\times& 
\exp i \Bigg[ 
L\int^T_{-T} dx^+\, j_1^c(x^+)\, j_2^c(x^+) + 
\frac{L+l}2 \int^t_{-t} dx^+\, \lp j_1^c(x^+)\, j_3^c(x^+) +
j_2^c(x^+)\, j_4^c(x^+)\rp 
\no \\
&+&
\frac{L-l}2 \int^t_{-t} dx^+\, \lp j_1^c(x^+)\, j_4^c(x^+)
+ j_2^c(x^+)\, j_3^c(x^+)\rp +
l\int^t_{-t} dx^+\, j_3^c(x^+)\, j_4^c(x^+) 
\Bigg]_{j_i=0}\, 
\eeeq
with $C_{\,1},\,\ldots\,,C_{\,4}$ as in Fig.~1 and $i=1,\,\ldots\,,4$.
Clearly, up to $\OO (g^4)$,
the only non-vanishing
contributions are those with an even number of derivatives both with
respect to $j_{1,2}$ and to $j_{3,4}$.

\begin{figure}[t]
\begin{center}
\epsfxsize=10cm
\epsffile{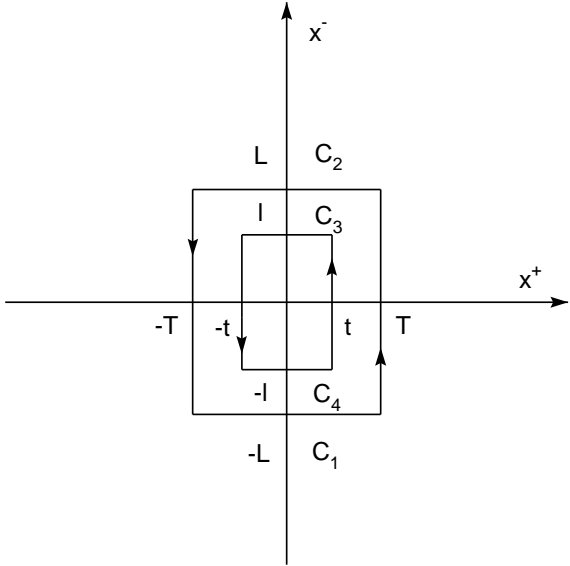}
\caption{Contours.}
\end{center}
\end{figure}

At $\OO (g^2)$, either a factor $iL$ or $il$ is produced when a
derivative acts on the first or the last term, respectively,  of the
last  exponential in Eq.~\re{variexp}, which represents the unique
contributions at this order, and finally $(2T)$ or $(2t)$, respectively
is given by integration over the loop variable. Thus we end up with
\beq\nome{quad}
\frac{i}4  g^2 \lp C_k A_2 + C_A A_1 \rp
\eeq
where $C_k=\frac{k (N-k) (N+1)}{N}$ and $C_A=2N$ are the quadratic
Casimir of the $k$-fundamental and of the adjoint representations,
respectively, and $A_2=4\,L\,T$, $A_1=4\,l\,t$ being the areas of the  
rectangles.
One can say, equivalently, the two loops factorize at this order, so
that, switching to the Euclidean space-time via a Wick rotation
and taking normalization  into account, $\wka^{({\rm II})}$ reads
\beq\nome{quadnor}
\wka^{({\rm II})}(A_1,A_2)= - g^2 \,\frac{C_A}4 \,A_1 \,.
\eeq

Likewise,   we can now easily derive the expression of
$\wka^{({\rm IV})}$, 
\ie the adjoint loop enclosed in
a  $k$-fundamental one  at  $\OO (g^4)$. 
Very schematically, different classes of diagrams can be  
distinguished. Let us browse on them.  Obvious contributions
are the pure $k$-fundamental loop  at $\OO (g^4)$ 
(which corresponds to $\tr \identity_{adj}$ in the inner loop) 
and the product of the pure adjoint by the pure $k$-fundamental loops  both at
$\OO (g^2)$, which turn out to result in 
\beq
\nome{factoriz}
-\frac{g^4}{32} \lp C_k^2\,A_2^2+2\,C_A\,C_k\,A_1\,A_2 \rp\,.
\eeq
On the other hand, it is clear  they will be removed by normalizing to
$\zk\wk$.
Very much in analogy with the first contribution, we have to consider
the pure adjoint loop at $\OO (g^4)$ (which corresponds to $\tr
\identity_k$ in the outer loop)
\beq
\nome{pure4}
-\frac{g^4}{32} \, C_A^2\,A_2^2 \,.
\eeq
However, the novelty in having one loop enclosed into another is
supplied by  graphs with propagators connecting sides of both
rectangles. We  naturally single out three prototypes, in which the
following situations occur
\begin{enumerate}
\item
both propagators are ``long'' (they travel a distance $L+l$); 
\item
both propagators are ``short'' (they travel a distance $L-l$)
\item
a propagator is ``short'' and the other one is ``long''.
\end{enumerate}
Pictorially, it is straightforward to recognize that there are three
diagrams falling within  both  the first  and  the second class, and
among them two graphs carry a factor $1/2$, owing to the presence of
integrals in the loop variables which are nested. Hence their
contribution to $\wka^{({\rm IV})}$ adds up to 
\beq
\nome{firstwo}
-\frac{g^4}4 \, t^2\,\frac{C_A\,C_k}{N^2-1}\lp \half + \half +1\rp 
\lq (L+l)^2+(L-l)^2 \rq\,.
\eeq
We emphasize that the factor $\frac{C_A\,C_k}{N^2-1}$ arises from
$\tr \lq t^at^b\rq_k \cdot \tr \lq T^aT^b\rq_{adj}$, properly normalized
to $(N^2-1){N \choose k}$ (the latter in $\zk\wk$).
Finally, four diagrams belong to the third class and yield
\beq 
\nome{third}
g^4 \, t^2\,\frac{C_A\,C_k}{N^2-1} (L^2-l^2)\,
\eeq
in which the opposite sign with respect to Eq.~\re{firstwo} has to be
ascribed  to the ``short''
propagator joining sides with the same orientation (the appearance of
the factor   
$\frac{C_A\,C_k}{N^2-1}$ has been explained above).

When the partial results Eqs.~(\ref{pure4}-\ref{third}) are summed up,
we obtain for $\wka$ in the Euclidean space-time
\beq
\nome{zit4}
\wka^{({\rm IV})}(A_1,A_2)=\frac{g^4}{32}  
\lp C_A^2\,A_1^2+ 4 \frac{C_A\,C_k}{N^2-1} \,A_1^2 \rp \,,
\eeq
which coincides with the $O(g^4)$ term  of Eq.~\re{wkdecfin}.
The perturbative LF formulation seems to capture again the exact
result, even in presence of a non-trivial topology.

\section{The instanton representation}

\noindent
As first pointed out by Witten \cite{witte}, one can represent the
partition  function ${\cal Z}_k$ in Eq.~(\ref{twist}) 
and the Wilson loop ${\cal W}_k$ in Eq.~(\ref{wilsonk}) on the sphere $S^2$ 
as sums over instanton contributions, following the procedure proposed
in \cite{gross1,gross} (see also \cite{case}). 
By instanton we mean a non-trivial classical solution of the
Yang-Mills  equations on $S^2$, which takes the form of an Abelian
Dirac  monopole embedded into the non-abelian gauge group.
For the general construction on any genus we refer to \cite{atiyah}.

Any of such configurations is characterized, in the $SU(N)$ or $U(N)$ case, 
by a set of $N$ integers $(n_1,\ldots,n_N)$. The set $(0,\ldots,0)$ represents
the topologically trivial solution. In Ref.~\cite{capo} it has been shown,
for the case of a Wilson loop in the fundamental representation
of the group $U(N)$, that it can be obtained by a {\it bona fide}
perturbative calculation \cite{tedeschi}
for the theory quantized in the light-cone
gauge by means of ET canonical commutators \cite{bosco,noi1}.

From the mathematical viewpoint, the zero-instanton sector
is reproduced  if  integration over the group  manifold
is replaced by integration over the tangent group algebra \cite{basso}.

Here we want to generalize such results to the case of a Wilson loop
in the adjoint representation for the $SU(N)$ theory, when $k$-sectors are 
taken into account. In so doing an intriguing interplay occurs
between instantons and $k$-states and one may wonder to what extent
perturbation theory can account for it.
In Sect.~IV we have shown that an adjoint loop in a $k$-vacuum state 
is equivalent to the same loop enclosed in a $k$-fundamental boundary one. 

In the following we will carry out the instanton expansion separately for the 
case of $SU(N)/Z_N$ in the $k$-sector ({\bf i}) and for the one with a
$k$-fundamental  boundary loop ({\bf ii}). We will find completely different
results. Let us begin with the former. 

\vspace{.5 truecm}

\noindent
({\bf i})

The instanton representation can be obtained by performing in 
Eqs.~(\ref{partip},\ref{wilsonint})
a Poisson resummation. Starting from Eq.~(\ref{partip}), it is 
convenient to introduce explicitly the constraint (\ref{defl}) in
a factorized form
\beeq
\nome{partifac}
&&\zk (A)=\frac{(2\pi)^{N-2}}{N!\,\sqrt\pi } \sum_{l_i=-\infty}^{+\infty} 
\int_{-\infty}^{+\infty}dl\,\int_0^{2 \pi}
d\a\, e^{-\lp \a - \frac{2\pi}N l \rp^2}
\de_{[N]} \lp k- l + \frac{N(N-1)}2 \rp \no \\
&&\ph{\zk (A)}
\times \int_{-\infty}^{+\infty}d\beta\,
\exp\Bigl[i\beta(l-\sum_{i=1}^{N}l_i)\Bigr]
\exp \left [ -\frac{g^2 A}{4} C_2 (l_i)\right ]
\Delta^2(l_1,...,l_N).
\eeeq
The Poisson transformation is a kind of duality relation
between two series
\begin{eqnarray}
\label{poisson}
&&\sum_{l_i=-\infty}^{+\infty}F(l_1,\ldots,l_N)=
\sum_{n_i=-\infty}^{+\infty}\tilde F(n_1,\ldots,n_N),\no \\
&&\tilde F(n_1,\ldots,n_N)=\int_{-\infty}^{+\infty}dz_1\ldots dz_N
F(z_1,\ldots,z_N)\exp\Bigl[2\pi i(z_1n_1+\ldots+z_Nn_N)\Bigr].
\end{eqnarray}
In order to perform the Fourier transform in (\ref{partifac}),
we remember that the transformation of a product is turned into
a convolution; moreover we recall the result 
\begin{eqnarray}
\label{ftrans}
&&\int_{-\infty}^{+\infty}dz_1\ldots dz_N
\exp\Bigl[i(z_1p_1+\ldots+z_Np_N)\Bigr]\Delta(\{z_i\})
\exp \lp -\frac{g^2A}{8}\sum_{q=1}^{N}z_q^2\rp= \no \\
&&\Bigl[\frac{4i}{g^2A}\Bigr]^{\frac{N(N-1)}{2}}\Bigl[\frac{8\pi}{g^2A}
\Bigr]^{\frac{N}{2}}\Delta(\{p_i\})
\exp\lp-\frac{2}{g^2A}\sum_{q=1}^{N}p_q^2\rp.
\end{eqnarray}
Taking these relations into account, Eq.~(\ref{partifac}) becomes
\begin{equation}
\label{sectors}
{\cal Z}_k(A)=\sum_{n=0}^{N-1}\exp\Bigl[\frac{2\pi ink}{N}\Bigr]
{\cal Z}^{(n)}(A),
\end{equation}
where
\begin{equation}
\label{nsector}
{\cal Z}^{(n)}(A)=(-1)^{n(N-1)}\,{\cal C}(A,N)\sum_{n_q=-\infty}^{+\infty}
\delta(n-\sum_{q=1}^{N}n_q)
\exp\Bigl[-\frac{4\pi^2}{g^2A}\sum_{q=1}^{N}(n_q-\frac{n}{N})^2\Bigr]
\zeta_n(\{n_q\})
\end{equation}
with $${\cal C}(A,N)=\frac{(2\pi)^{2N-3}}{N!}\sqrt{\pi N} \,e^{\frac{g^2A}{48}
N(N^2-1)} \,2^{N(N+\frac{1}{2})}\, (g^2A)^{-(N^2-N/2-1/2)}$$ and
\begin{equation}
\label{zeta}
\zeta_n(\{n_q\})=\int_{-\infty}^{+\infty}dz_1\ldots dz_N \exp\lq-
\frac{1}{2}\sum_{q=1}^{N} z_q^2\rq \Delta ( \{
\sqrt{\frac{g^2A}{2}} z_q+2 \pi n_q\} )
\Delta(\{\sqrt{\frac{g^2A}{2}}z_q-2\pi 
n_q\}).
\end{equation}
Eqs.~(\ref{sectors},\ref{nsector}) exactly provide the explicit form of the 
partition function in the $k$-sector in the instanton representation, 
as anticipated in Eqs.~(\ref{sector2},\ref{angle2}). ${\cal Z}^{(n)}(A)$ is 
the contribution from the $n$-th topological 
sector and  is localized around the classical solutions of the Yang-Mills 
equations in that sector. 
According to the localization picture, in 
Eq.~(\ref{nsector}) we can readily single out  the contribution of the  
classical instanton action 
$\exp\Bigl[-\frac{4\pi^2}{g^2A}\sum_{q=1}^{N}(n_q-\frac{n}{N})^2\Bigr]$: 
it is remarkable 
that the non-trivial bundle structure ($n\neq 0$) induces a shift in the 
instanton numbers from integral to fractional quantities, 
$n_q\to n_q-\frac{n}{N}$, while the 
$\delta$-constraint  properly implements the tracelessness condition for a 
$SU(N)$ matrix. 
Moreover, $\zeta_n(\{n_q\})$ represents the contribution of the quantum 
fluctuations 
around the classical solutions. We notice that the coefficient $C(A,N)$ is 
singular as $\sqrt{g^2 A}\rightarrow 0$: this is expected because
zero-modes  appear when computing fluctuations in the instanton
background, the total degree of singularity depending on the instanton
numbers, as the polynomial part in Eq.~\re{zeta} shows \cite{gross1}. 
The only non-exponentially suppressed contribution, in that limit, comes from 
the $n=0$ sector, as argued by Witten in \cite{witte}, and, in particular, only
the fluctuations around the trivial connections survive.

The same procedure is now to be performed for ${\cal W}_k(A_1,A_2)$
starting from Eq.~(\ref{wilsonint}). We obtain the instanton representation
\begin{equation}
\label{wsectors}
{\cal W}_k(A_1,A_2)=\frac{1}{N+1}+\frac{1}{{\cal Z}_k}(A)\sum_{n=0}^{N-1}
\exp\Bigl[\frac{2\pi ink}{N}\Bigr]
{\cal W}^{(n)}(A_1,A_2)
\end{equation}
where
\beeq
\label{nwilson}
&&{\cal W}^{(n)}=(-1)^{n(N-1)}\, \frac{2\,{\cal C}(A,N)}{N^2-1}
\exp\Big[\frac{g^2(A_1-A_2)^2}{8A}\Big]\sum_{r<s}
\sum_{n_q=-\infty}^{+\infty}\delta(n-\sum_{q=1}^{N}n_q)
\no \\
&&\ph{{\cal W}^{(n)}} \times 
\exp \lq-\frac{4\pi^2}{g^2A}\sum_{q=1}^{N}(n_q-\frac{n}{N})^2 \rq
\exp \lq 2\pi i(n_s-n_r)\frac{A_2}{A}\rq\Omega_n(\{n_q\})
\eeeq
and
\beeq
\label{fluct}
&&\Omega_n(\{n_q\})=\int_{-\infty}^{+\infty}dz_1\ldots dz_N
\exp\lq -\frac{1}{2}\sum_{q=1}^{N}z_q^2\rq \exp\lq\frac{i}{2}
\sqrt{\frac{g^2A}{2}}(z_r-z_s)\rq
\times \no \\
&&\Delta(\sqrt{\frac{g^2A}{2}}z_1-\tilde n_1,\ldots,
\sqrt{\frac{g^2A}{2}}z_N-\tilde n_N)\,
\Delta(\sqrt{\frac{g^2A}{2}}z_1+\tilde n_1,\ldots,
\sqrt{\frac{g^2A}{2}}z_N+\tilde n_N)\,,
\eeeq
with
\beq
\label{tilde}
\tilde n_q=2\pi n_q-(\delta_{q,r}-\delta_{q,s})\,\frac{ig^2(A_1-A_2)}{4}\,.
\eeq
Following Ref.~\cite{gross}, in Eq.~(\ref{nwilson}) 
one can still single out 
the classical instanton actions and their classical 
contributions to
the Wilson loop $\lp\exp \lq 2\pi i(n_s-n_r)\frac{A_2}{A}\rq \rp$.

We now focus our attention on the zero-instanton
sector ($n_q=0$, $\forall$ $q$) 
with the purpose of exploring its relation with possible
perturbative treatments.

Taking symmetry under permutations into account, 
Eqs.~(\ref{sectors}-\ref{zeta}) and (\ref{wsectors}-\ref{fluct}) become
\beq
\label{zetazero}
{\cal Z}_k^{(0)}(A)={\cal C}(A,N)
\int_{-\infty}^{+\infty}dz_1\ldots dz_N \exp\lq-
\frac{1}{2}\sum_q z_q^2\rq \Delta^2(\sqrt{\frac{g^2A}{2}}\{z_q\})
\eeq
and
\begin{eqnarray}
\label{wzero}
&&{\cal W}_k^{(0)}(A_1,A_2)=\frac{1}{N+1}\Bigg[1+\frac{{\cal C}(A,N)}
{{\cal Z}_k^{(0)}(A)}\exp\Big[\frac{g^2(A_1-A_2)^2}{8A}\Big]\times \no \\
&&\int_{-\infty}^{+\infty}dz_1\ldots dz_N
\exp\lq -\frac{1}{2} \sum_{q=1}^{N}z_q^2\rq \exp\lq\frac{i}{2}
\sqrt{\frac{g^2A}{2}}(z_1-z_2)\rq
\times \no \\
&&\Delta(\sqrt{\frac{g^2A}{2}}z_1+\frac{ig^2}{4}(A_1-A_2),\sqrt{\frac{g^2A}{2}}
z_2-\frac{ig^2}{4}(A_1-A_2),\sqrt{\frac{g^2A}{2}}z_3,
\ldots,\sqrt{\frac{g^2A}{2}}z_N)\times \no \\
&&\Delta(\sqrt{\frac{g^2A}{2}}z_1-\frac{ig^2}{4}(A_1-A_2),\sqrt{\frac{g^2A}{2}}z_2+
\frac{ig^2}{4}(A_1-A_2),\sqrt{\frac{g^2A}{2}}z_3,
\ldots,\sqrt{\frac{g^2A}{2}}z_N)\Bigg].
\eeeq
We remark that the dependence on $k$ has completely disappeared; the label $k$
will  be thereby dropped.

We
express the Vandermonde determinants in terms of Hermite
polynomials and then expand them using the completely
antisymmetric tensor. Afterwards, we integrate over $z_3,\ldots,z_N$, taking 
orthogonality into account, and are left with the expression
\begin{eqnarray}
\label{deter}
&&\varepsilon^{j_1,j_2,j_3,\ldots,j_N}\varepsilon_{q_1,q_2,j_3,\ldots,j_N}
(-1)^{j_2-q_2}\times \nonumber \\
&&\int_{-\infty}^{+\infty}dz_1
\exp\left[-\frac{1}{2}
z_{1}^2\right]
\exp\left(\frac{i\sqrt {g^2A}\,z_{1}}
{2\sqrt 2}\right)He_{j_1}(z_{1+})He_{q_1}(z_{1-})\times \nonumber \\
&&\int_{-\infty}^{+\infty}dz_2
\exp\left[-\frac{1}{2}
z_{2}^2\right]
\exp\left(\frac{i\sqrt {g^2A}\,z_{2}}
{2\sqrt 2}\right)He_{j_2}(z_{2+})He_{q_2}(z_{2-}),
\end{eqnarray}
where
\begin{equation}
\label{zetapm}
z_{1,2\pm}=z_{1,2}\pm\frac{i}{4} \sqrt{\frac{2g^2}{A}}(A-2A_1).
\end{equation}

We have now the remarkable relation \cite{basso}
\begin{eqnarray}
\label{remarkable}
&&\int_{-\infty}^{+\infty}dz_1
\exp\left[-\frac{1}{2}
z_{1}^2\right]
\exp\left(\frac{i\sqrt {g^2A}\,z_{1}}
{2\sqrt 2}\right)He_{q}(z_{1+})He_{r}(z_{1-})=\nonumber \\
&&\exp \lq  -\frac{g^2}{16A} \lp A-2A_1 \rp^2 \rq
\left(A-A_1\right)^{\frac{q-r}{2}}A_1^{\frac{r-q}{2}}
\times\nonumber \\
&&\int_{-\infty}^{+\infty}dz_1\exp\left[-\frac{1}{2}
z_{1}^2\right]
\exp \lp igz_1\sqrt{\frac{A_1(A-A_1)}{2A}} \rp He_{q}(z_1)He_{r}(z_1). \no
\end{eqnarray}
Thanks to it, Eq.~(\ref{wzero}) takes the form
\begin{eqnarray}
\label{wilzero}
&&{\cal W}^{(0)}(A_1,A_2)=\frac{1}{N+1}+\frac{N}{{\cal Z}
\,(N+1)}\int_{-\infty}^{+\infty}dz_1\ldots dz_N
\exp \lq -\frac{1}{2}\sum_{j=1}^{N}z_j^2 \rq \times\nonumber \\
&&\exp\Bigl[ig(z_1-z_2)
\sqrt\frac{A_1(A-A_1)}{2A}\Bigr] \Delta^2(z_1,\ldots,z_N)=  
\nonumber \\
&&\frac{1}{{\cal Z}}\int {\cal D}F\,
\exp(-\half {\rm Tr}F^2)\frac{1}{N^2-1} \lq \abs {{\rm Tr} \lp
\exp(igF{\cal E})\rp }^2-1 \rq,
\end{eqnarray}
where ${\cal Z}=\int {\cal D}F\,
\exp(-\frac{1}{2}{\rm Tr}F^2).$

Here ${\cal E}=
\sqrt{\frac{A_1(A-A_1)}{2A}}$ 
and ${\cal D}F$ denotes the flat integration measure 
on the tangent space of constant Hermitian traceless $N\times N$ matrices. 
The restriction to traceless matrices  only affects the
overall normalization in this case.
The powers appearing in Eq.~(\ref{remarkable})
cancel owing to the presence of the antisymmetric tensor. 

Retaining only the zero-instanton sector is therefore equivalent to 
integrating over the group algebra \cite{basso}. On the other hand, if in 
Eq.~(\ref{wilzero}) we perform the decompactification limit $A\to \infty$,
we exactly recover the perturbative result in which we have used the WML
propagator \cite{tedeschi,capo,basso}. 
Memory of the $k$-th topological sector has been completely
lost. 

\vspace{.5 truecm}

\noindent
({\bf ii})

We now address the issue of singling the zero-instanton (trivial) sector out
for the adjoint loop enclosed in a $k$-fundamental one. 
The result we are going to obtain will keep a $k$-dependence and 
will by no means correspond 
to any perturbative calculation, at variance  with the preceding case.
One should not be too surprised by this conclusion; as a matter of fact,
the instanton structures of the two cases, as long as one remains 
on the sphere, are completely different. Only in the decompactification
limit, when all instantons are summed in both cases, the
same limit ensues; but there is no reason why this miracle should occur
when the two (different) zero-instanton sectors are compared.

The calculations that follow, although conceptually simple, are 
rather heavy and we shall try to condense them as much as possible.
Starting from Eq.~\re{wilsonkhar},
we define $$\tilde n_1=2\pi n_1-i\frac{g^2}{4}(2A_3-A), \, \ldots\,,
\,\tilde n_k=2\pi n_k-i\frac{g^2}{4}(2A_3-A),\,\tilde n_{k+1}=
2\pi n_{k+1},\,\ldots\,,\,\tilde n_N=2\pi n_N.$$
After a long calculation we are led to the result
\beeq
\label{wilpois}
&&\wk\, {\cal Z}_k=\sum_{n_q=-\infty}^{+\infty}\,\delta \lp \sum_{q=1}^{N} n_q\rp
\, \exp\lq\frac{g^2k^2A_3(A-A_3)}{4NA}\rq
\exp\lq\frac{g^2k}{16A}(2A_3-A)^2\rq \no \\
&&\exp \lq -\frac{4\pi^2}{g^2A}\sum_{q=1}^{N} n_q^2 \rq\exp\lp \frac{2 i\pi A_3}{A}
\sum_{q=1}^kn_q \rp
\int_{-\infty}^{+\infty}dy_1\ldots dy_N 
\exp \lq -\frac{1}{g^2A}\sum_{q=1}^{N} y_q^2 \rq \no \\
&&\exp\Big(-\frac{i}{2}\sum_{q=1}^ky_q\Big)\,
\Delta(y_j+\tilde n_j)\Delta(y_j-\tilde n_j).
\eeeq
We remark the quite different structure of the classical instanton
action when compared to the one of the preceding section Eq.~\re{wzero}.

The zero-instanton sector is  obtained again by choosing $n_q=0$, $\forall$ 
$q$. Eq.~(\ref{wilpois}) becomes
\beeq
\label{wilkzero}
&&\wk^{(0)}\, {\cal Z}_k^{(0)}=
\exp\lq\frac{g^2k^2A_3(A-A_3)}{4NA}\rq
\exp\lq\frac{g^2k}{16A}(2A_3-A)^2\rq\times \no \\
&&\int_{-\infty}^{+\infty}dy_1\ldots dy_N 
\exp\lq-\frac{1}{g^2A}\sum_{q=1}^{N} y_q^2\rq 
\exp\lp-\frac{i}{2}\sum_{q=1}^ky_q\rp\times \no \\
&&\Delta (y_1-i\frac{g^2}{4}(2A_3-A),\ldots,y_k-i\frac{g^2}{4}(2A_3-A),
y_{k+1},\ldots,y_N)\no \\
&&\Delta (y_1+i\frac{g^2}{4}(2A_3-A),\ldots,
y_k+i\frac{g^2}{4}(2A_3-A),y_{k+1},\ldots,y_N).
\eeeq
Now, rescaling the variables, expanding the Vandermonde determinants
in terms  of Hermite
polynomials and taking Eq.~(\ref{remarkable}) into account, 
Eq.~(\ref{wilkzero}) assumes the form
\beeq
\label{wilkwzero}
&&\wk^{(0)}\, {\cal Z}_k^{(0)}=
\exp\lq\frac{g^2k^2A_3(A-A_3)}{4NA}\rq
\int_{-\infty}^{+\infty}dz_1\ldots dz_N 
\exp\Big[-\frac{1}{2}\sum_{q=1}^{N} z_q^2\Big] 
\times \no \\
&&\Delta^2(z_1,\ldots,z_n)\, \exp\lq ig\sqrt{\frac{A_3 (A-A_3)}
{2A}}\sum_{q=1}^k z_q\rq=\no \\
&&\exp\lq\frac{g^2k^2A_3(A-A_3)}{4NA}\rq\int\,{\cal D} F
\exp(-\frac{1}{2}\tr F^2)\tr \lq \exp\Big(igF\sqrt{\frac{A_3(A-A_3)}
{2A}}\Big)\, \rq_k, 
\eeeq
F being a constant 
$N\times N$ hermitian matrix and the trace being taken in the
$k$-fundamental
representation. The exponential factor is needed to
turn the $U(N)$ representation into one of $SU(N)$.
As a matter of fact, the previous equation can be written as
\beq
\label{tracel}
\wk^{(0)}\, {\cal Z}_k^{(0)}=
\int\,{\cal D} F
\exp(-\frac{1}{2}\tr F^2) \tr \lq \exp\Big(igF\sqrt{\frac{A_3(A-A_3)}
{2A}}\Big)\, \rq_k,
\eeq
with a {\it traceless} matrix $F$.
Again keeping the zero-instanton sector is equivalent to 
integrating over the group algebra.

The instanton expansion of $\wka$ is rather cumbersome and we present
here just a brief account. Referring to Eq.~\re{corre}, four cases are
now possible:
\begin{itemize}
\item $q_1<q_2\le k$ with weight ${k\choose 2},$
\item $k<q_1<q_2$ with weight ${N-k\choose 2},$
\item $q_1\le k < q_2$ with weight $\frac{k(N-k)}{2},$
\item $q_2\le k <q_1$ with weight  $\frac{k(N-k)}{2}.$
\end{itemize}
Correspondingly, the second term in the r.h.s. of Eq.~(\ref{corre})
splits into  four contributions
\beeq
\label{corres}
&&{\cal W}_{k,a}(A_1,A_2,A_3)=\frac{1}{N+1}+\frac{2}{{\cal Z}_k
{\cal W}_k\,(N^2-1)}{N\choose k}\sum_{l_j}
\int_{0}^{2\pi}d\alpha \exp \Big[-(\alpha-\frac{2\pi}{N}l)^2\Big]
\times \no\\
&& \exp\Big[-\frac{g^2A}{4}C(l_j)-\frac{g^2}{2}A_1\Big]
\exp\Big[-\frac{g^2A_3}{4}\Big(\frac{k(N-k)}{N}-2\sum_{j=1}^k(l_j-\frac{l}{N})
\Big)\Big] \Delta(l_j^R)
\times \no \\
&&\Big[{k\choose 2} \,\Delta_{(1)}(l_j^T)\,e^{\frac{g^2A_1}{2}(l_1-l_2)}+
{N-k\choose 2} \,\Delta_{(2)}(l_j^T)\,e^{\frac{g^2A_1}{2}(l_{N-1}-l_N)}
+ \no \\
&&\frac{k(N-k)}{2}
\Big( \Delta_{(3)}(l_j^T) \,e^{\frac{g^2A_1}{2}(l_1-l_N)}
+\Delta_{(4)}(l_j^T)\,e^{-\frac{g^2A_1}{2}(l_1-l_N)}\Big)\Big]=\no \\
&&\frac{1}{N+1}+\frac{2}{{\cal Z}_k {\cal W}_k\,(N^2-1)}{N\choose k}
\lq \wka^{(1)}+\wka^{(2)}+\wka^{(3)}+\wka^{(4)}\rq \,,
\eeeq
where $\Delta_{(i)}(l_j^T)$, $i=1,\ldots,4$, is the Vandermonde
determinant in the four cases above and $\wka^{(i)}(A_1,A_2,A_3)$ is 
conveniently defined.
At this stage, each of the $\wka^{(i)}$'s has to undergo the same
treatment of $\wk \zk$ (Eq.~\re{wilpois}). For $\wka^{(i)}$ we define
\beq
\label{ntilde}
\tilde{n}^{(i)}_j=2\pi n_j +h^{(i)}_j\,
\eeq
with $h^{(i)}_j$ collected in Table~I.
\begin{table}[h]
\begin{center}
\caption{Values of $h^{(i)}_j$, defined in Eq.~\re{ntilde}. The indices
$i$ and $j$  label the columns and the rows, respectively.}
\begin{tabular}{|c||c|c|c|c|}
$h^{(i)}_j$&1&2&3&4\\
\hline
1&$i\frac{g^2}{2}A_2$&$i\frac{g^2}{4}(A-2A_3)$&
$i\frac{g^2}{2}A_2$&
$-i\frac{g^2}{2}(A_3-A_1)$\\
\hline
2&$i\frac{g^2}{2}(A_1-A_3)$&
$i\frac{g^2}{4}(A-2A_3)$ &
$i\frac{g^2}{4}(A-2A_3)$&
$i\frac{g^2}{4}(A-2A_3)$\\
\hline
$3,\ldots,k$&$i\frac{g^2}{4}(A-2A_3)$&
$i\frac{g^2}{4}(A-2A_3)$&
$i\frac{g^2}{4}(A-2A_3)$&
$i\frac{g^2}{4}(A-2A_3)$\\
\hline
$k+1,\ldots,N-2$&
$0$&$0$&$0$&$0$\\
\hline
$N-1$&
$0$&
$i\frac{g^2}{4}(A-2A_1)$&
$0$&
$0$\\
\hline
$N$&$0$&
$-i\frac{g^2}{4}(A-2A_1)$&
$-i\frac{g^2}{4}(A-2A_1)$&
$i\frac{g^2}{4}(A-2A_1)$\\
\end{tabular} 
\end{center}
\end{table}
After that, the Poisson transform for  $\wka^{(i)}$ reads
\beeq
\label{wilipois}
&&\wka^{(i)}=\sum_{n_q=-\infty}^{+\infty}\,\delta \lp \sum_{q=1}^{N} n_q\rp
\, \exp\lq-\frac{g^2A_1}2 \lp 1-\frac{A_1}A\rp\rq
\exp\lq-\frac{g^2A_3}2 \lp 1-\frac{A_3}A\rp\frac{k(N-k)}{2N}\rq\no \\
&&\exp \lq -\frac{4\pi^2}{g^2A}\sum_{q=1}^{N} n_q^2 \rq\exp\lq \frac{2 i\pi}{A}
\lp A_3 \sum_{q=1}^kn_q + M^{(i)}(n_j;A_1,A_2,A_3)\rp \rq  \\
&&\int_{-\infty}^{+\infty}dy_1\ldots dy_N 
\exp \lq -\frac{1}{g^2A}\sum_{q=1}^{N} y_q^2 \rq 
\exp\frac{i}{2}\lp 
Y^{(i)}(y_j)+\sum_{q=2}^ky_q\rp\,
\Delta(y_j+\tilde n_j^{(i)})\Delta(y_j-\tilde n_j^{(i)})\,,\no
\eeeq
where
\beq
\label{defM}
M^{(i)}(n_j;A_1,A_2,A_3)=
 \left\{
\begin{array}{ll}
A_1(n_1-n_2) & \qquad i=1\\
A_1(n_{N-1}-n_N) & \qquad i=2\\
-(A_2 +A_3 )n_1 -A_1 n_N & \qquad i=3\\
A_1 (n_N -n_1)& \qquad i=4
\end{array}
\right.
\eeq
and
\beq
\label{defyp}
Y^{(i)}(y_j)= \left\{
\begin{array}{ll}
y_2 & \qquad i=1\\
y_1-y_{N-1}+y_N & \qquad i=2\\
y_N & \qquad i=3\\
2y_1-y_N & \qquad i=4\,.
\end{array}
\right.
\eeq
Choosing  $n_q=0$, $\forall q$, in Eq.~\re{wilipois} and inserting  it
in \re{corres}, we obtain the zero-instanton sector
\beeq
\label{zero2loop}
&&{\cal W}_{k,a}^{(0)}(A_1, A_2, A_3)= \frac1{N+1} + 
\frac2{(N^2-1) }
\, \exp\lq {-\frac{g^2A_1}2}\lp 1- \frac{A_1}A \rp \rq \frac1{\hat{I}}\times  
\\
&&\lq {k \choose 2} \hat{I}_1 + {N- k \choose 2} \hat{I}_2 + 
\frac{k(N-k)}2  \, \exp \lp \frac{g^2}2 \frac{A_3 A_1}A \rp \hat{I}_3 
+ \frac{k(N-k)}2 \exp \lp \frac{-g^2}2 \frac{A_3 A_1}A\rp \hat{I}_4 \rq\,, \no
\eeeq
where   $\hat{I}$ is the integral over $z_j$ appearing in
Eq.~\re{wilkwzero} and $\hat{I}_i(A_3, A_2, A_1)$, $i=1, \ldots, 4$ are
explicitly given in the Appendix.
We are interested in the limit $A_3 \to \infty$, $A_2$, $A_1$ fixed of
${\cal W}_{k,a}^{(0)}$, which, far from being trivial, reads
\beeq
\label{2loop}
&&{\cal W}_{k,a}^{(0)}(A_1,A_2,A_3 \to \infty)= \frac1{N+1} + 
\frac2{(N^2-1) I}
\, e^{\frac{-g^2A_1}2} \times \no \\
&&\lq {k \choose 2} I_1 + {N- k \choose 2} I_2 + 
\frac{k(N-k)}2  \, e^{\frac{g^2A_1}2} I_3 
+ \frac{k(N-k)}2 e^{\frac{-g^2A_1}2} I_4 \rq\,.
\eeeq
For the sake of brevity, we defer the computation of
$I=\hat{I}(A_3 \to \infty, A-A_3)$ and
$I_i=\hat{I}_i(A_1, A_2, A_3 \to \infty)$ to  the Appendix and report
here just the results
\beeq
\label{defintlim}
I&=&\int_{-\infty}^{+\infty}
dz_1\ldots dz_N \exp\Big[-\frac{1}{2}\sum_{q=1}^{N} z_q^2\Big]
H(z_1,\ldots,z_N;A_1,A_2)\,,\no \\
I_i&=&\int_{-\infty}^{+\infty}
dz_1\ldots dz_N \exp\Big[-\frac{1}{2}\sum_{q=1}^{N} z_q^2\Big]
H_i(z_1,\ldots,z_N;A_1,A_2)\,,
\eeeq
where
\beeq
\label{varint}
&&H=\Delta ( z_1+1,\ldots,z_k+1,z_{k+1}\ldots,
z_N)
\Delta ( z_1-\frac{g^2}2 \hat{A},\ldots,z_k-\frac{g^2}2 \hat{A},
z_{k+1}\ldots,z_N)\,,\no \\
&&H_1=\Delta\lp z_1,z_2+2,
z_3+1,\ldots,z_k+1,z_{k+1}\ldots,
z_N\rp \times  \no \\ 
&&\Delta\lp z_1-\frac{g^2}2 A_2,z_2-\frac{g^2}2 (\hat{A}+A_1), 
z_3-\frac{g^2}2 \hat{A},\ldots,z_k-\frac{g^2}2\hat{A},z_{k+1}\ldots,
z_N\rp\,,\no \\
&&H_2=\Delta\lp z_1+1,\ldots,z_k+1,
z_k+1,z_{k+1}\ldots,
z_{N-2}, z_{N-1}-1,z_N+1\rp\times \no \\ 
&&\Delta\lp z_1-\frac{g^2}2 \hat{A},\ldots,z_k-\frac{g^2}2 \hat{A},
z_{k+1}\ldots,z_{N-2},z_{N-1}+\frac{g^2}2A_1,z_N-\frac{g^2}2A_1\rp \,,\no \\
&&H_3=\Delta\lp z_1,z_2+1,\ldots,z_k+1,
z_{k+1},\ldots,z_{N-1},z_N+1\rp\times \no \\ 
&&\Delta\lp z_1-\frac{g^2}2 A_2,z_2-\frac{g^2}2\hat{A},\ldots,
z_k-\frac{g^2}2 \hat{A},
z_{k+1}\ldots,z_{N-1},z_N-\frac{g^2}2A_1\rp\,, \no \\
&&H_4=\Delta\lp z_1+2,z_2+1,\ldots,z_k+1,
z_{k+1},\ldots,z_{N-1},z_N-1\rp
\times \no \\ 
&&\Delta\lp z_1-\frac{g^2}2 (\hat{A}+A_1),z_2-\frac{g^2}2
\hat{A},\ldots,
z_k-\frac{g^2}2 \hat{A},
z_{k+1}\ldots,z_{N-1},z_N+\frac{g^2}2A_1\rp \,, 
\eeeq 
with $\hat{A}=A_1+A_2$.

In order to compare the zero-instanton sectors of the single adjoint
loop and of the two-loop correlators, the additional limit $A_2 \to
\infty$, $A_1$ fixed has to be performed in \re{2loop}. The leading
contribution in $A_2$ of $I$, $I_i$ is straightly proven to be
$A_2^{k(N-k)}$,
so that it cancels out in  \re{2loop} and the limit is
finite. Unfortunately, we were not successful in computing it exactly
for all values of $N$ and $k$,
and resolved to restrict ourselves to simple cases. We carried the
calculations out for $N=2,3,4$, with all possible value of $k$
($k=0,\ldots,N-1$) and found  
\begin{itemize}
\item $N=2$, $k=0$ 
$$
\WW_{0,a}^{(0)}(A_1)=\frac13\lq 1+2\,e^{-\frac{g^2 A_1}2}\lp 1-g^2 A_1 \rp \rq 
\,;
$$
\item $N=2$, $k=1$ 
$$
\WW_{1,a}^{(0)}(A_1)=e^{-g^2 A_1}\,;
$$
\item  $N=3$, $k=0$
$$
\WW_{0,a}^{(0)}(A_1)=\frac1{4} \lq 1+4 \,e^{-\frac{g^2 A_1}2}
\lp 12-18g^2 A_1+\frac92 g^4 A_1^2 -\frac12 g^6 A_1^3  \rp \rq\,;
$$
\item  $N=3$, $k=1$ (or, equivalently, $k=2$)
$$
\WW_{1,a}^{(0)}(A_1)=\frac1{8} \lq 1-4 e^{-\frac{g^2 A_1}2}+ \lp 11-3g^2 A_1\rp
e^{-g^2 A_1}\rq\,;
$$
\item    
$N=4$, $k=0$
\beeqn
\WW_{0,a}^{(0)}(A_1)=\frac1{15} \Bigg[ 3
+ 
&\Big( & 12-24g^2 A_1+\frac{23}2g^4 A_1^2 \\
&-&\frac83g^6 A_1^3
+\frac{25}{96}g^8 A_1^3-\frac1{96}g^{10} A_1^5\Big)
e^{-\frac{g^2 A_1}2}\Bigg]\,;
\eeeqn 
\item    
$N=4$, $k=1$ (or, equivalently, $k=3$)
\beeqn
\WW_{1,a}^{(0)}(A_1)=\frac1{30} \Big[ 4
&+& \lp -36 +18g^2 A_1 -3g^4 A_1^2\rp
e^{-\frac{g^2 A_1}2}\\
&+&\lp 62 -34 g^2 A_1+ 3g^4 A_1^2 \rp e^{-g^2 A_1}\Big]\,;
\eeeqn
\item 
$N=4$, $k=2$
$$
\WW_{2,a}^{(0)}(A_1)=\frac1{15} \lq 1+ \lp 14 -16g^2 A_1 + 3 g^4 A_1^2 \rp
e^{-g^2 A_1} \rq\,.
$$
\end{itemize} 
We notice the following basic features:
\begin{enumerate}
\item
for $k \neq 0$, the results are different from the single loop case;
\item 
for $k=0$, the limit $A_2 \to \infty$ is immaterial, as
can be understood  from Eq.~\re{varint}.
Actually, the case at hand corresponds to taking
the identical representation sitting in the outer loop, so that the
correlator becomes insensitive to $A_2$ being finite or infinite;
\item
although string tensions are independent of $k$, as Eq.~\re{2loop}
explicitly shows, the polynomial coefficients do  depend on it.
\end{enumerate}

\section{Perturbative series defined via WML prescription}

We now check that the zero instanton contribution to 
the expectation value of an adjoint loop enclosed in
a  $k$-fundamental one  on the plane, expressed by
Eq.~\re{2loop},  is consistent, at least up to $O(g^4)$, with the
perturbative computation where the propagator is prescribed according
to WML.

The starting point is again  the perturbative definition  of $\wka$ in
the light-cone gauge Eq.~\re{wcurr}. 
At variance with Sect.~V, the
propagator $D(x-y)$ is now given by Eq.~\re{WMLprop}~\footnote{Henceforth, 
for notational simplicity, we will adopt the same symbol $\wka$ for the 
perturbative WML two-loop correlator.}. 
The choice of circles for the contours $\G_k$, $\G_a$ in the Euclidean
space-time (we recall the
invariance under area-preserving diffeomorphisms) will prove  
particularly convenient  \cite{tedeschi}; 
the currents  will be accordingly  defined
to have  support on the contours.

The weighted basic correlator
\beq
\label{basic}
\dot{x}_-(s) \,\dot{x}_-(s') \, \frac{x_+ (s)- x_+ (s')}{x_- (s) -
x_-(s')}\,,
\eeq
turns out to be  independent of the loop variables
when describing  a propagator which starts and ends on the same
contour, 
and amounts to $2(\pi r)^2=2\pi A_1$ for the inner circle  and to
$2(\pi R)^2=2\pi A_2$ for the outer circle. After that, for a diagram
containing only  propagators of that kind, integration
over the path parameters is
trivial and one is left with
the purely combinatorial problem of determining the group factors.
At $\OO (g^2)$ the two loops factorize, so that we end up with
\beq\nome{quadml}
-\frac{g^2}4   \lp C_k A_2 + C_A A_1 \rp
\eeq
When normalization is taken into account, $\wka^{({\rm II})}$ reads
\beq\nome{quadnorml}
\wka^{({\rm II})}(A_1,A_2)= -\frac{g^2A_1}4\,C_A 
\eeq
and coincides with Eq.~\re{quadnor} as expected \cite{noi1}.

Likewise,   we can now easily derive the expression of
$\wka^{({\rm IV})}$, 
\ie the adjoint loop enclosed in
a $k$-fundamental one  at  $\OO (g^4)$. 

Very schematically, different classes of diagrams can be  
distinguished. Let us browse on them.  Obvious contributions
are the pure $k$-fundamental loop  at $\OO (g^4)$ 
(which corresponds to $\tr \identity_{adj}$ in the inner loop) 
and the product of the pure adjoint by the pure $k$-fundamental loops  both at
$\OO (g^2)$, which turn out to result in 
\beq
\nome{factorizml}
\frac{g^4}{16}\lq \frac{A_1^2}2 \lp C_k^2-\frac16 \,C_A\,C_k \rp +
A_1\,A_2 \,C_A\,C_k \rq\,.
\eeq
On the other hand, it is clear  they will be removed by normalizing to
$\zk\wk$.
Very much in analogy with the first contribution, we have to consider
the pure adjoint loop at $\OO (g^4)$ (which corresponds to $\tr
\identity_k$ in the outer loop)
\beq
\nome{pure4ml}
\frac5{192}\, g^4 \, C_A^2\,A_1^2 \,.
\eeq

However, the novelty in having one loop enclosed in another is
supplied by  graphs with propagators joining  the two circles. 
In this case, the weighted basic correlator can be inferred from
Eq.~\re{basic} and reads
\beq
\nome{bas2}
2\pi^2 r R \, \frac{\frac{r}R e^{2\pi i s}-e^{2\pi i s'}}{e^{2\pi i s}-
\frac{r}R e^{2\pi i s'}}\,.
\eeq
Integration over the loop variables (with $r<R$) at $\OO (g^4)$   and
insertion of the proper group factors yield   the contribution of
those graphs to $\wka^{({\rm IV})}$ 
\beq
\nome{inout}
\frac{g^4}8 \, \frac{C_A\,C_k}{N^2-1} A_1^2\,.
\eeq
We emphasize that the factor $\frac{C_A\,C_k}{N^2-1}$ arises from
$\tr \lq t^at^b\rq_k \cdot \tr \lq T^aT^b\rq_{adj}$, properly normalized
to $(N^2-1){N \choose k}$ (the latter in $\zk\wk$).

When the partial results Eqs.~(\ref{pure4ml},\ref{inout}) are summed up,
we obtain for $\wka$ 
\beq
\nome{zit4ml}
\wka^{({\rm IV})}(A_2,A_1)=\frac{g^4A_1^2}8  
\lp \frac5{24}\,C_A^2+ \frac{C_A\,C_k}{N^2-1} \rp\,.
\eeq

At this point a comment is in order. The coincidence of the
coefficients of $g^4A_1^2 \frac{C_A\,C_k}{N^2-1}$ in Eqs.~\re{zit4} and
\re{zit4ml}   should not be too surprising: in fact, they correspond
to graphs connecting the inner adjoint loop to the outer
$k$-fundamental one, in the LF and ET formulation, respectively, and
at this order those graphs are in some sense ``abelian-like'', since
no crossing  takes place (equivalently, the traces in the adjoint and
in the $k$-fundamental representation are trivial). 
In fact, the different behaviour of CPV and WML
prescription in such diagrams is expected to arise only at
$\OO(g^6)$, when  at least two propagators  out of three, or more,
joining the two loops, cross. The appearance of a dependence on $A_2$
is also to be ascribed to  graphs with crossing propagators. 

As we have already announced in the previous section, we were not
successful in treating the limit $A_2\to \infty$ in
Eqs.~(\ref{2loop}-\ref{varint}) for all values of $N$ and
$k$. Nevertheless, we  argue that in such a limit the zero-instanton
contribution  of the two-loop correlator has to coincide with the
perturbative series defined above.
Eqs.~(\ref{defintlim},\ref{varint}) suggest for
$\WW_{k,a}^{(0)}(A_1,A_2)$ the following form
\beq
\label{argu}
\WW_{k,a}^{(0)}(A_1,\hat{A})=1-\frac{C_A}4\,  g^2A_1 +\sum_{n=2}  
(g^{2}A_1)^n\,R_n(g^2\hat{A})\,,
\eeq
where $R_n(g^2\hat{A})=\frac{Q_n(g^2\hat{A})}{P_n(g^2\hat{A})}$, both $Q_n$
and $P_n$ being polynomials of degree $k(N-k)$.
In Eq.~\re{argu} the dependence on $A_2$ has been conveniently
rearranged into $\hat{A}$, which is the natural variable in the
normalization factor $I$ (see \re{wilkwzero}). One should be careful
not to regard Eq.~\re{argu} as an expansion in $g^2$, as the rational
functions $R_n$ produce a series in $g^2$ of their own. Nonetheless,
such a form emphasizes the finiteness of the limit $\hat{A} \to \infty$, $A_1$
fixed.
Notice that the contributions
$\OO(g^6)$ contain the whole dependence on $A_2$, which is naturally  
expected to appear in the perturbative expansion, as remarked
beforehand.
Moreover, Eq.~\re{argu} points out one has simply to set $\hat{A}=0$ to
recover  the contribution $\OO(g^4)$. This we did for a limited sample 
of values of $N$ ($N=2,3,4$ with all possible values of $k$) and found
the terms up to $\OO(g^4)$ in
the expansion of Eq.~\re{2loop} are  reproduced by
Eqs.~(\ref{quadnorml},\ref{zit4ml}).

We stress the only case in which the WML perturbative expansion
turns out  to be independent of $A_2$ is $k=0$; in fact, it coincides
with the expansion inferred from the exact results for the two-loop
correlator reported in Sect.~VI. This is not at all surprising, since
Eq.~\re{zit4ml} and the expression of $C_k$ appearing in Sect.~V point
out  that, at least $\OO (g^4)$, the correlator
reduces,  for  $k=0$, to a single adjoint loop of area
$A_1$. Nonetheless, the argument can be pushed further and it is easy
to verify it holds at any order.

\section{Conclusions}

\noindent
The main goal of this paper was the study of vacua for $SU(N)/Z_N$ gauge
theories in two dimensions, trying to generalize previously obtained results
\cite{capo,noi} concerning a Wilson loop in the fundamental representation,
to the adjoint case.

The motivation was twofold: on one hand matter in the adjoint 
representation (infinitely heavy fermions in our case) can mimic the effects
of ``transverse'' degrees of freedom, which are obviously lacking in
our two-dimensional world; on the other hand, as long as only adjoint 
representations are involved, the true symmetry group becomes $SU(N)/Z_N$,
those representations being insensitive to the group center.
As a result, the topological properties of the theory are modified.

This feature induces indeed a much richer 
topological structure: many inequivalent vacua occur, which are the 
non-abelian counterpart of the familiar $\theta$-vacua of the Schwinger model
(or of $QCD_4$).

If the theory is considered on a sphere $S^2$ with area $A$ (which is
eventually to be decompactified sending $A\to \infty$),  in 
the $SU(N)$ case already an infinite set of topological 
excitations (instantons)
are present \cite{witte}. 
They are responsible for the big difference we found between 
an ET and a LF description of the theory: confinement at large $N$, which 
can be easily obtained in the LF formulation, can  be only recovered in the
ET scenario if those instantons are fully taken into account.
This situation strengthens the belief that the LF vacuum provides
indeed a simpler picture, at least in two dimensions.

In the $SU(N)/Z_N$ case, the presence of a non-trivial bundle structure
leads to inequivalent $k$-vacua and deeply modifies the instanton 
pattern on $S^2$. Nevertheless, we found that a LF vacuum is again closer 
to the exact solution, which can be reached by summing a 
suitable perturbative series.
Moreover we have shown that different $k$-sectors of the theory can be 
interpreted as due to the presence of $k$-charges at $\infty$ in the form
of a boundary Wilson loop in the $k$-fundamental representation, as
conjectured by Witten long ago \cite{wittheta}.

However, this property holds only for the exact solution and in the 
decompactification limit; it is not shared by the zero-instanton contribution
which corresponds to the perturbative ET result.
We find remarkable that a simple, yet deep feature of the theory
emerges only after all nonperturbative effects are taken into account.

At this stage, we think we have set a solid ground for the most
interesting  future development, namely the introduction of dynamical
fermions, with a particular focus on the generation and on the
properties of  a chiral condensate.
This problem has already been tackled in the recent literature
\cite{fugle}, without reaching so far firm conclusions in the  
non-supersymmetric case.

\section*{Acknowledgements}

One of us (AB) acknowledges useful discussions with A. Zhitnitsky,
which stimulated his interest on this topic. He also
wishes to thank B. McKellar for the hospitality he enjoyed at the
School of Physics of the University of Melbourne, where part of this work 
was done. This work was carried out in the framework of the NATO project
``QCD Vacuum Structure and Early Universe'' granted under the reference
PST.CLG 974745.

\section{Appendix}

\noindent
We report here the integrals $\hat{I}_i(A_1,A_2,A_3)$, $i=1,\ldots,4$,
making their appearance in Eq.~\re{zero2loop}
\beeq 
\label{defint}
&&\hat{I}_1=\int_{-\infty}^{+\infty}
dz_1\ldots dz_N 
\exp\Big[-\frac{1}{2}\sum_{q=1}^{N} z_q^2\Big]\times  \\ 
&&\Delta\lp
z_1-ig\frac{A_2}{\sqrt{2A}},z_2+ig\frac{A+A_3-A_1}{\sqrt{2A}},
z_3+ig\frac{A_3}{\sqrt{2A}},\ldots,z_k+ig\frac{A_3}{\sqrt{2A}},z_{k+1}\ldots,
z_N\rp\times            \no            \\                  &&\Delta\lp
z_1+ig\frac{A_2}{\sqrt{2A}},z_2+ig\frac{A-A_3+A_1}{\sqrt{2A}},
z_3+ig\frac{A-A_3}{\sqrt{2A}},\ldots,z_k+ig\frac{A-A_3}{\sqrt{2A}},z_{k+1}\ldots,
z_N\rp\,, \no \\ &&\hat{I}_2= \int_{-\infty}^{+\infty} dz_1\ldots dz_N
\exp\Big[-\frac{1}{2}\sum_{q=1}^{N} z_q^2\Big]\times \no \\ 
&&\Delta\lp z_1+ig\frac{A-A_3}{\sqrt{2A}}, \ldots,
z_k+ig\frac{A-A_3}{\sqrt{2A}},
z_{k+1}\ldots,z_{N-2},
z_{N-1}-ig\frac{A_1}{\sqrt{2A}},
z_N+ig\frac{A_1}{\sqrt{2A}}\rp\times \no \\
&&\Delta\lp z_1+ig\frac{A_3}{\sqrt{2A}}, \ldots,
z_k+ig\frac{A_3}{\sqrt{2A}},
z_{k+1}\ldots,z_{N-2},
z_{N-1}-ig\frac{A-A_1}{\sqrt{2A}},
z_N+ig\frac{A-A_1}{\sqrt{2A}}\rp\,, \no \\
&&\hat{I}_3= \int_{-\infty}^{+\infty}
dz_1\ldots dz_N 
\exp\Big[-\frac{1}{2}\sum_{q=1}^{N} z_q^2\Big]\times \no \\ 
&&\Delta\lp z_1+ig\frac{A_2}{\sqrt{2A}}, z_2+ig\frac{A-A_3}{\sqrt{2A}}, 
\ldots,
z_k+ig\frac{A-A_3}{\sqrt{2A}},
z_{k+1}\ldots,z_{N-1},
z_N+ig\frac{A_1}{\sqrt{2A}}\rp\times \no \\
&&\Delta\lp z_1-ig\frac{A_2}{\sqrt{2A}}, z_2+ig\frac{A_3}{\sqrt{2A}}, 
\ldots,
z_k+ig\frac{A_3}{\sqrt{2A}},
z_{k+1}\ldots,z_{N-1},
z_N+ig\frac{A-A_1}{\sqrt{2A}}\rp \,,\no \\
&&\hat{I}_4= \int_{-\infty}^{+\infty}
dz_1\ldots dz_N 
\exp\Big[-\frac{1}{2}\sum_{q=1}^{N} z_q^2\Big]\times \no \\ 
&&
\Delta\lp z_1+ig\frac{A+A_3-A_1}{\sqrt{2A}}, z_2+ig\frac{A_3}{\sqrt{2A}}, 
\ldots,
z_k+ig\frac{A_3}{\sqrt{2A}},
z_{k+1}\ldots,z_{N-1},
z_N-ig\frac{A-A_1}{\sqrt{2A}}\rp \times \no \\
&&
\Delta\lp z_1+ig\frac{A-A_3+A_1}{\sqrt{2A}}, z_2+ig\frac{A-A_3}{\sqrt{2A}}, 
\ldots,
z_k+ig\frac{A-A_3}{\sqrt{2A}},
z_{k+1}\ldots,z_{N-1},
z_N-ig\frac{A_1}{\sqrt{2A}}\rp  \,. \no
\eeeq 
As announced in Sect.~VI, we want to show how the limit $A_3\to\infty$
can be performed so as to obtain $I(A_1,A_2)$, $I_i(A_1,A_2)$. We
address $\hat{I}_1$ as an example; the other cases can be dealt with
the same technology.
The product of the two Vandermonde in the integral over
$z_1,\ldots,z_N$, with Gaussian measure, can be rewritten in terms of
generalized Laguerre  polynomials as follows \cite{erd}
\beeq
\label{laguerre}
&&(2\pi)^{N/2} \prod_{\a=k+1}^N  (n_{\a}-1)!\,
\e^{j_1\,\ldots \,j_k \,j_{k+1}\,\ldots \,j_N}\,
\e^{l_1\, \ldots \, l_k \, j_{k+1}\, \ldots \, j_N}\,
\Bigg\{ 
(j_1-1)!\lp ig \frac{A_2}{\sqrt{2A}}\rp^{l_1-j_1}
\no\\
&& L^{(l_1-j_1)}_{j_1-1}\lp -\frac{g^2 A_2^2}{2A}\rp 
\Bigg\} \Bigg\{
(j_2-1)!\lp ig \frac{A-A_3+A_1}{\sqrt{2A}}\rp^{l_2-j_2}
L^{(l_2-j_2)}_{j_2-1}\lp g^2\frac{ A^2-(A_3-A_1)^2}{2A}\rp 
\Bigg\} 
\no\\ &&
\prod_{q=3}^k  
\Bigg\{ 
(j_q-1)!\lp ig \frac{A-A_3}{\sqrt{2A}}\rp^{l_q-j_q}
L^{(l_q-j_q)}_{j_q-1}\lp g^2\frac{ A_3(A-A_3)}{2A}\rp 
\Bigg\}\,.
\eeeq
The next step consists in factorizing 
all possibly divergent terms out of Eq.~\re{laguerre}, which, in the limit 
$A_3\to\infty$ ($A\to\infty$),  amount to $(ig \sqrt{A/2})^{\sum_{q=1}^k
(j_q-l_q)}$. 
Now, recalling that
$\sum_{q=1}^k j_q=\sum_{q=1}^k l_q$, we are left with the finite expression
\beeq
\label{laguerre1}
&&(2\pi)^{N/2} \prod_{\a=k+1}^N  (n_{\a}-1)!\,
\e^{j_1\,\ldots \,j_k \,j_{k+1}\,\ldots \,j_N}\,
\e^{l_1\, \ldots \, l_k \, j_{k+1}\, \ldots \, j_N}\no \\
&&
\Bigg\{(j_1-1)!\lp - \frac{g^2 A_2}{2}\rp^{l_1-j_1}
L^{(l_1-j_1)}_{j_1-1}\lp 0\rp  \Bigg\}\no \\
&&\Bigg\{ (j_2-1)!
\lp -g^2 \frac{2A_1+A_2}{2}\rp^{l_2-j_2}
L^{(l_2-j_2)}_{j_2-1}\lp g^2 ( 2A_1+A_2)\rp 
\Bigg\} \no\\
&&\prod_{q=3}^k  
\Bigg\{ (j_q-1)!\lp -g^2 \frac{A_1+A_2}{2}\rp^{l_q-j_q}
L^{(l_q-j_q)}_{j_q-1}\lp g^2 \frac{A_1+A_2}{2}\rp 
\Bigg\}\,,
\eeeq
which can be recombined into
\beeq
\label{laguerre2}
&&\Delta\lp z_1,z_2+2,
z_3+1,\ldots,z_k+1,z_{k+1}\ldots,
z_N\rp \times   \\ 
&&\Delta\lp z_1-\frac{g^2}2 A_2,z_2-\frac{g^2}2 (A_2+2A_1),
z_3-\frac{g^2}2(A_2+A_1),\ldots,z_k-\frac{g^2}2(A_2+A_1),z_{k+1}\ldots,
z_N\rp\,. \no
\eeeq
This is precisely the integrand $H_1$ in Eqs.~(\ref{defintlim},\ref{varint}).

\vfill\eject
\end{document}